\documentclass[superscriptaddress,aps,prd,reprint]{revtex4-2}

\usepackage{lipsum}
\usepackage{blindtext}
\usepackage{graphicx}
\usepackage{tensor}
\usepackage{dcolumn}
\usepackage{bm}

\usepackage{hyperref}
\usepackage{slashed}
\usepackage{amsfonts}
\usepackage{mathrsfs}
\usepackage{color}
\usepackage[makeroom]{cancel}
\usepackage{graphicx}
\usepackage{subcaption}
\usepackage{amssymb}
\usepackage{amsmath}
\usepackage{hyperref}
\usepackage{cleveref}

\def\bi{\begin{itemize}}
\def\ei{\end{itemize}}
\def\be{\begin{equation}}
\def\ee{\end{equation}}
\newcommand{\bea}{\begin{eqnarray}}
\newcommand{\eea}{\end{eqnarray}}

\begin{document}

\begin{flushright} {\footnotesize YITP-26-38}  \end{flushright}

\allowdisplaybreaks[4]
\preprint{APS/123-QED}
\title{A Unified Fermionic Origin for Density-Driven Waterfall Inflation and its Multi-Messenger Signatures}

\author{Stephon~Alexander}
\affiliation{Brown Theoretical Physics Center and Department of Physics,Brown University,RI 02903, USA}

\author{Pisin~Chen}
\affiliation{Leung Center for Cosmology and Particle Astrophysics, National Taiwan University, Taipei 10617, Taiwan}
\affiliation{Department of Physics and Graduate Institute of Astrophysics, National Taiwan University, Taipei 10617, Taiwan}
\affiliation{Kavli Institute for Particle Astrophysics and Cosmology, SLAC National Accelerator Laboratory, Stanford University, CA 94305, U.S.A.}

\author{Jinglong~Liu}
\affiliation{Tsung-Dao Lee Institute, 1 Lisuo Road, Shanghai, 201210, China}
\affiliation{School of Physics and Astronomy, Shanghai Jiao Tong University, 800 Dongchuan Road, Shanghai 200240, China}

\author{Antonino~Marcian\`o}
\affiliation{Center for Field Theory and Particle Physics \& Department of Physics, Fudan University, 200433 Shanghai, China}
\affiliation{Center for Astronomy and Astrophysics, Fudan University, 200433 Shanghai, China}
\affiliation{Laboratori Nazionali di Frascati INFN, Frascati (Rome), Italy, EU}
\affiliation{INFN sezione Roma Tor Vergata, I-00133 Rome, Italy, EU}

\author{Misao~Sasaki}
\affiliation{Asia Pacific Center for Theoretical Physics, Pohang 37673, Korea}
\affiliation{Kavli Institute for the Physics and Mathematics of the Universe (WPI), The University of Tokyo Institutes for Advanced Study, The University of Tokyo, Chiba 277-8583, Japan}
\affiliation{Center for Gravitational Physics, Yukawa Institute for Theoretical Physics, Kyoto University, Kyoto 606-8502, Japan}
\affiliation{Leung Center for Cosmology and Particle Astrophysics, National Taiwan University, Taipei 10617, Taiwan}

\author{Xuan-Lin~Su}
\affiliation{Center for Field Theory and Particle Physics \& Department of Physics, Fudan University, 200433 Shanghai, China}
\affiliation{Center for Astronomy and Astrophysics, Fudan University, 200433 Shanghai, China}










\begin{abstract}
\noindent 
Inflation is conventionally described in terms of fundamental scalar fields whose microscopic origin remains unknown. We propose a fundamentally different mechanism in which inflation emerges from fermion condensates generated by torsion-induced four-fermion interactions in Einstein--Cartan--Holst gravity. Starting from a purely fermionic theory, we derive a two-condensate effective description whose inflationary dynamics resembles hybrid inflation but exhibits a qualitatively new exit mechanism. 
The central result is the emergence of a \emph{density-driven waterfall transition}. During inflation, axial charge production continuously increases the fermion density and the associated chiral chemical potential. Solving the finite-density gap equation, we show that above a critical axial density the condensate no longer admits a symmetry-broken solution. Inflation therefore terminates through the evaporation of the condensate itself rather than through the inflaton crossing a critical field value. This mechanism naturally triggers rapid energy transfer to fermionic degrees of freedom and provides a microscopic realization of instant preheating. 
The post-transition condensate is unstable to fragmentation into charged non-topological solitons. These objects can subsequently collapse into primordial black holes, generating a calculable relic population whose properties are determined by the same condensate dynamics that drives inflation. Because the underlying theory is intrinsically chiral and connected to dynamical Chern--Simons gravity, the scenario simultaneously predicts parity-violating signatures.
The framework unifies inflation, reheating, soliton formation, primordial-black-hole production, and parity-violating phenomenology within a single gravitationally induced fermion-condensation mechanism. It provides a composite alternative to scalar-field inflation and identifies correlated observational signatures that can test the role of spacetime torsion in the early Universe.
\end{abstract}
\maketitle

\section{Introduction}
\noindent 
The inflationary paradigm provides the most successful description of the very early Universe, explaining the observed homogeneity, isotropy, and spatial flatness of cosmological spacetime while simultaneously generating the primordial fluctuations that seeded the formation of large-scale structures \cite{Guth:1980zm,LINDE1983177,Albrecht:1982wi}. Despite its remarkable phenomenological success, the microscopic origin of inflation remains unknown. In most realizations, inflation is driven by one or more fundamental scalar fields whose particle-physics origin is left unspecified. This raises the question of whether inflation may instead emerge from more fundamental degrees of freedom already present in nature.

An attractive possibility is that the relevant scalar degrees of freedom are not elementary but composite. This idea finds support in several areas of physics, ranging from superconductivity and superfluidity to dynamical symmetry breaking in quantum field theory. In particular, four-fermion interactions can generate fermion condensates through mechanisms analogous to the Bardeen-Cooper-Schrieffer (BCS) instability and the Nambu-Jona-Lasinio (NJL) model. Such condensates naturally behave as effective scalar degrees of freedom at low energies and may therefore provide a dynamical origin for inflation \cite{Inagaki:1993ya,Tong:2023krn,Alexander:2009uu,Shapiro:2001rz,alexander2009cosmological,Addazi:2017qus}.

Gravity itself provides a natural framework for generating these interactions. In Einstein-Cartan theory and its Holst extension, fermions source spacetime torsion, which can be integrated out to yield effective four-fermion interactions \cite{Mercuri:2006um,Alexander:2014eva,Alexander:2014uaa,Freidel:2005sn}. When the resulting interaction is attractive, a fermion condensate may form, generating an effective scalar sector entirely from gravitational dynamics. In this perspective, inflation is not driven by a fundamental inflaton field but emerges from collective fermionic degrees of freedom generated by spacetime geometry itself.

A fermion-condensate realization of inflation was previously proposed in Ref.~\cite{Addazi:2017qus}, where the effective potential generated by the condensate was shown to support a slow-roll phase consistent with cosmological observations. More recently, Ref.~\cite{Alexander:2025whu} suggested that multiple condensate sectors may generate a structure analogous to hybrid inflation \cite{Linde:1993cn}. While these studies established the viability of condensate-driven inflation, several fundamental questions remained open. In particular, a complete microscopic description of the end of inflation, reheating, fragmentation, and the possible generation of compact relics was still lacking.

In this work we present a substantially extended framework in which these phenomena arise from a common dynamical origin. The key observation is that the fermion condensate is intrinsically sensitive to finite-density effects. During inflation, the production of chiral fermions continuously increases the axial charge density and the associated chiral chemical potential. Solving the finite-density gap equation, we show that above a critical density the condensate ceases to admit a symmetry-broken solution. Inflation therefore terminates through the evaporation of the condensate itself. Unlike conventional hybrid inflation, where the waterfall transition is triggered when a scalar field crosses a critical value, the transition studied here is controlled by the accumulation of fermionic density.

This mechanism constitutes the central result of the present work. We refer to it as a \emph{density-driven waterfall transition}. The transition simultaneously provides a graceful exit from inflation and a microscopic realization of instant preheating. Because the energy stored in the condensate is directly coupled to the underlying fermionic sector, the collapse of the condensate rapidly transfers energy into fermionic degrees of freedom and initiates the radiation-dominated epoch.

The dynamical waterfall has further consequences. The post-transition condensate is unstable to fragmentation into non-topological solitons. In the present framework these objects arise naturally as Q-balls associated with the composite condensate sector \cite{Coleman:1985ki,kusenko1997small,Kusenko:1997si,Kusenko:2005du}. Such solitons have long been known to play an important role in cosmology \cite{Enqvist:1998en,Kasuya:2001hg,Kasuya:2014ofa,Cotner:2019ykd}, and under suitable conditions may collapse into primordial black holes (PBHs) \cite{Cotner:2016cvr,Cotner:2016dhw,Cotner:2017tir,Cotner:2019ykd}. Consequently, the same mechanism that ends inflation can also generate a relic population of compact objects that may contribute to the dark matter abundance \cite{Zeldovich:1967lct,hawking1971gravitationally,carr1974black,1975ApJ...201....1C,Garcia-Bellido:1996mdl,Khlopov:2008qy,Frampton:2010sw,Kawasaki:2016pql,Carr:2016drx,Inomata:2016rbd,Inomata:2017okj,Georg:2017mqk,Chen:2002tu,Chen:2004ft,Adler:2001vs,Sasaki:2016jop}.

A further motivation for studying this framework is that the underlying fermionic interactions are closely connected to parity-violating gravitational dynamics. Self-interacting fermions naturally induce dynamical Chern–Simons gravity \cite{Alexander:2022cow}, leading to characteristic parity-violating effects in cosmology, such as the propagation and production of gravitational waves. These include birefringence and damping effects \cite{Liu:2025ifb}, thereby providing observational probes of the same microscopic physics responsible for inflation. The framework therefore predicts correlated signatures across several observational channels, including primordial black holes, stochastic gravitational-wave backgrounds, parity-violating tensor perturbations, and reheating observables.

The picture that emerges is remarkably economical. Starting from gravity coupled to fermions, torsion generates attractive four-fermion interactions; these interactions produce composite condensates; the condensates drive inflation; finite-density effects trigger a dynamical waterfall transition; the subsequent fragmentation generates solitonic relics and primordial black holes; and the same chiral structure predicts parity-violating gravitational-wave signatures. Inflation, reheating, compact-object formation, and phenomenology thus arise from a common underlying mechanism rooted in spacetime torsion.

\subsection*{Central Claim}
\noindent 
The principal results established in this work are summarized as follows:

\begin{itemize}
\item \textbf{Condensate-driven Hybrid Inflation:} We derive a two-condensate effective theory from torsion-induced four-fermion interactions in Einstein--Cartan--Holst gravity, realizing hybrid inflation without introducing fundamental scalar fields.
\item \textbf{Density-driven Waterfall Mechanism:} We identify a critical finite-density threshold that destroys the condensate solution of the gap equation, providing a microscopic mechanism to naturally terminate inflation.
\item \textbf{Instant Preheating:} We show that the condensate's collapse rapidly transfers its vacuum energy into fermionic degrees of freedom, yielding a natural realization of instant preheating.
\item \textbf{Soliton and PBH Formation:} We demonstrate that the post-transition condensate is unstable against fragmentation into non-topological solitons (Q-balls), which can subsequently seed primordial-black-hole (PBH) formation.
\item \textbf{Parity-Violation Signatures:} We connect the underlying chiral fermionic sector to parity-violating signatures, in particular, the gravitational-wave observables from the effective Chern-Simons gravity, linking the inflationary dynamics and astrophysical phenomenology within a single unified framework.
\end{itemize}

The remainder of this paper is organized as follows. Section II reviews the emergence of torsion-induced four-fermion interactions in Einstein--Cartan--Holst gravity. Section III constructs the composite condensate description and the resulting effective inflationary dynamics. Section IV introduces the finite-density gap equation and the density-driven waterfall transition. Section V details the rapid energy transfer to fermionic degrees of freedom and the reheating dynamics. Sections VI and VII investigate the formation of Q-balls and primordial black holes, respectively. Section VIII analyzes the associated parity-violating gravitational-wave signatures. We conclude in Section IX with a discussion of the broader implications of this framework.

\section{Overview of the Density-Driven Waterfall Mechanism}
\label{sec:overview}
\noindent 
This section outlines the physical foundation and dynamical evolution of fermion condensate hybrid inflation. By unifying spacetime torsion, fermion self-interactions, and finite-density effects, this framework provides a natural mechanism for inflation, a graceful density-driven exit, and instantaneous preheating, ultimately setting the stage for the formation of primordial black holes.

\subsection{ECH Gravity and the NJL Model}
\noindent 
The foundation of this model rests on the Einstein-Cartan-Holst (ECH) action, which extends General Relativity by admitting spacetime torsion. Unlike metric gravity, torsion couples non-minimally to fermions. Because torsion is non-dynamical at low energies, the contorsion tensor can be integrated out, generating an effective four-fermion interaction. The resulting interaction is characterized by the contraction of two axial currents, $J_5^L = \overline{\psi}\gamma_5\gamma^L\psi$, namely
\begin{equation}
S_{int} = -\xi\kappa^2 \int_M d^4x |e| J_5^L J_5^M \eta_{LM}.
\end{equation}

By applying the Fierz identity and the Pauli-Kofink relation, the vector terms are removed, simplifying the interaction into an attractive combination of scalar and pseudoscalar bilinears. 

Because the prefactor $-\xi\kappa^2$ can be large and negative, it provides a strong, attractive self-interaction. In analogy to the Nambu-Jona-Lasinio (NJL) model of quantum chromodynamics, this interaction triggers a Bardeen-Cooper-Schrieffer (BCS) style condensation of fermions in the early universe. This macroscopic condensate breaks chiral symmetry and generates a Coleman-Weinberg-like effective potential capable of driving slow-roll inflation.

\begin{figure}[htbp]
\centering
\includegraphics[width=0.45\textwidth]{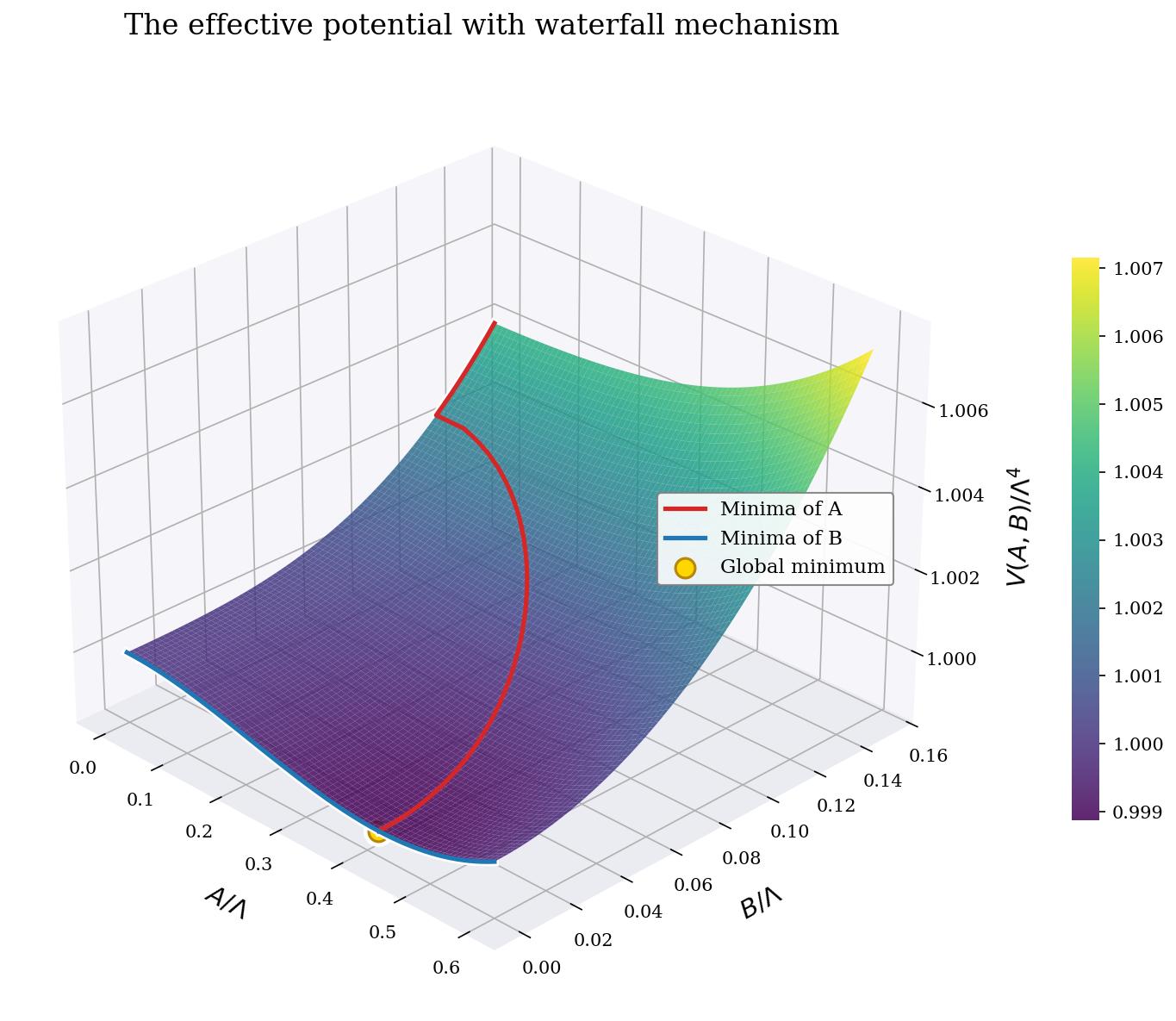}
\caption{Surface plot of the normalized effective potential $V(A,B)/\Lambda^4$. The $B$-axis is truncated at $0.15\Lambda$ to clearly resolve the field-space topography. The field-dependent minima for the waterfall field $A$ and the inflaton field $B$ are traced by the red and blue curves, respectively. For this illustrative parameter choice ($m^2 / \Lambda^2 = 1 / 2\pi^2$), the auxiliary gap field $A$ undergoes a sharp phase transition as the slowly rolling inflaton approaches its origin. The global minimum of the potential, marked by the gold point, lies at the intersection of these two minimum trajectories, visually confirming the dynamical waterfall mechanism.}
\label{fig:Potential}
\end{figure}

To realize a hybrid inflationary scenario, we partition the fermion field into two distinct sectors, $\psi$ and $\chi$. Through a Hubbard-Stratonovich transformation, we integrate out these fermions to introduce two composite bound states (gap fields), $A$ and $B$:
\begin{itemize}
    \item \textbf{The Inflaton Field ($B$):} The bound state $B$ serves as the slowly rolling inflaton.
    \item \textbf{The Waterfall Field ($A$):} The field $A$, associated with the fermion condensate, acts as the auxiliary waterfall field.
\end{itemize}
The resulting effective potential governing this hybrid-like inflationary scenario takes the form
\begin{equation}
V(A,B) = V_0 + \frac{m^2}{2}A^2 - m_\phi^2 |B|^2 + \tilde{V}_{\rm NJL}(A,B).
\end{equation}
Here, $\tilde{V}_{\rm NJL}(A,B)$ represents the standard Nambu-Jona-Lasinio potential \cite{Inagaki:1993ya, Addazi:2017qus}, generalized by the substitution $A \rightarrow A + yB$. The parameter $y$ is an effective coupling constant generated at the one-loop level upon integrating out the second fermionic sector \cite{Alexander:2022cow}. Crucially, the dynamical running of the inflaton field $B$ can trigger a phase transition in the gap field $A$, naturally realizing a waterfall mechanism. This symmetry-breaking trajectory is illustrated in Fig.~\ref{fig:Potential}.

\subsection{Density-Driven Waterfall: Dynamics and Intuition}
\noindent 
In standard hybrid inflation, the exit is triggered geometrically when the slowly rolling inflaton crosses a critical threshold in field space, destabilizing the waterfall field. In contrast, the exit mechanism in the fermion condensate model is intrinsically thermodynamic and governed by the temporal accumulation of axial charge. 

The introduction of an axial chemical potential $\mu$ breaks the $U(1)_A$ symmetry, preventing the conservation of the Noether current and allowing for continuous net fermion number production as the universe expands. During inflation, this production continuously increases the chemical potential $\mu(t)$, which functions as a Fermi energy, gradually building up a dense Fermi sea of real fermions. 

The state of the condensate is dictated by the gap equation, $\partial V(A)/\partial A = 0$. In the presence of the chemical potential, the modified gap equation explicitly reflects the available phase space, i.e. 
\begin{equation}
\begin{aligned}
\label{eq:gap}
\frac{1}{\lambda} = \frac{1}{\pi^3} \int_0^\Lambda |\vec{q}|^2 d|\vec{q}| &\left[ \frac{1}{E_+} \arctan\left(\frac{W_q}{E_+}\right)\right. \\
&\left. + \frac{1}{E_-} \arctan\left(\frac{W_q}{E_-}\right) \right],   
\end{aligned}
\end{equation}
where $E_{\pm} = \sqrt{(|\vec{q}| \pm \mu)^2 + A^2}$ and $W_q \equiv \sqrt{\Lambda^2 - |\vec{q}|^2}$. 
\begin{figure}
\centering
\includegraphics[width=0.4\textwidth]{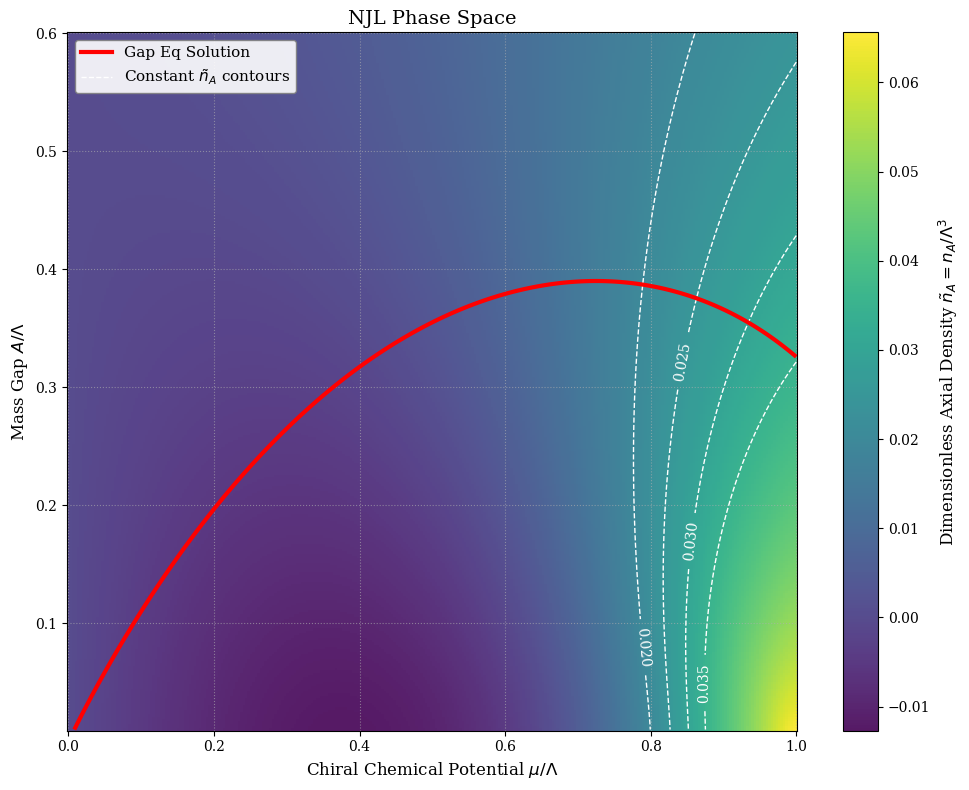}
\caption{Phase space evaluation of the chiral condensate. The background color gradient maps the magnitude of the right-hand side of the chemical potential equation of motion. Dashed white contours represent constant values of the normalized axial number density, $\langle n_A \rangle/\Lambda^3$. The solid red curve traces the real-valued solutions to the gap equation (the vacuum manifold). Physical states for a given number density are localized at the intersections between the constant-density contours and the gap equation curve. Here, the curvature parameter is set to $m^2 / \Lambda^2= 1/\pi^2$.}
\label{fig:condensateEOM}
\end{figure}

Concurrently, the temporal evolution of the axial number density $\langle n_A \rangle$ is captured by the chiral anomaly expression evaluated in a Friedmann-Lema\^{i}tre-Robertson-Walker (FLRW) background, namely 
\begin{equation}
\label{eq:anomaly}
\partial_\tau \langle n_A \rangle + 3\mathcal{H}\langle n_A \rangle = \frac{2A_0^2 a}{\lambda}.
\end{equation}

The physical mechanism for the inflationary exit is driven directly by the interplay of these two equations via \textbf{Pauli blocking}. As the universe evolves, the density term in Eq.~(\ref{eq:anomaly}) grows proportionately with conformal time ($\propto \tau^2$), increasing $\mu$. This growing Fermi sea exerts degeneracy pressure. Looking at Eq.~(\ref{eq:gap}), as $\mu$ grows, the denominator $E_{\pm}$ increases, suppressing the integral. Physically, the real fermions occupy states that would otherwise be available, invoking strict Pauli blocking that prevents the virtual fermions in the Dirac sea from pairing into the coherent state required to sustain the macroscopic condensate $A$. 

Once the chemical potential hits a critical limit ($\mu \sim \Lambda$), the integral in Eq.~(\ref{eq:gap}) can no longer satisfy the $1/\lambda$ threshold. The gap equation loses its real-valued solution. The energetic cost of maintaining the broken symmetry outpaces that of a free Fermi gas, causing the condensate to instantaneously ``melt'' ($A \rightarrow 0$) and triggering a density-driven waterfall transition.

\subsection{Instant Preheating and Soliton Formation}
\noindent 
The destruction of the false vacuum instigates a violent tachyonic instability, rapidly transferring the vacuum energy into the kinetic energy of the gap field $A$. Because the waterfall field $A$ is fundamentally a composite fermion condensate, it inherently possesses an $\mathcal{O}(1)$ effective Yukawa coupling to the underlying fermions. Consequently, the explosive energy of the scalar fluctuations is instantly dumped back into the pre-existing Fermi sea. 

Driven by the strong four-fermion interactions intrinsic to the NJL framework, the system rapidly thermalizes. This ensures an instantaneous preheating phase and a seamless transition to a radiation-dominated era with a high reheating temperature $T_{rh} \sim \Lambda$. 

During the waterfall phase, the effective potential's second derivative exhibits negative regions, meaning the post-transition condensate is generically unstable against fragmentation. This tachyonic growth halts only when the field stabilizes at larger values, leading to the formation of localized, non-topological solitons --- specifically, Q-balls. Because these stable configurations carry a conserved global charge, they can dominate the energy density and, under appropriate gravitational conditions, undergo collapse. This mechanism ultimately acts as the seeding process for primordial black holes (PBHs), linking the early-universe fermion dynamics directly to dark matter abundance.

\section{Torsion-Induced Four-Fermion Theory}
\label{sec:ech}
\noindent 
The microscopic origin of the present framework lies in the coupling
between fermionic matter and spacetime torsion. In Einstein--Cartan
gravity, fermions do not couple exclusively to the metric geometry,
but also source the antisymmetric part of the affine connection through
their intrinsic spin density
\cite{hehl1973spin,hehl1974spin,lord1976tensors,
de1994spin,Shapiro:2001rz,hammond2002torsion}.
A convenient starting point is provided by the
Einstein--Cartan--Holst (ECH) action
\cite{Mercuri:2006um,Alexander:2014eva,Alexander:2014uaa},
which combines the gravitational and fermionic sectors,
\begin{equation}
S_{\rm ECH}
=
S_{\rm GR}
+
S_{\rm Dirac}.
\end{equation}
The gravitational part is
\begin{equation}
S_{\rm GR}
=
\frac{1}{2\kappa^2}
\int d^4x\, |e|\,
e^\mu_I e^\nu_J
P^{IJ}{}_{KL}
R^{KL}{}_{\mu\nu},
\end{equation}
where
\begin{equation}
P^{IJ}{}_{KL}
=
\delta^{[I}_K\delta^{J]}_L
-
\frac{1}{2\gamma}
\epsilon^{IJ}{}_{KL},
\end{equation}
and $\gamma$ denotes the Barbero--Immirzi parameter.
The fermionic sector is described by
\begin{equation}
S_{\rm Dirac}
=
\frac12
\int d^4x\, |e|\,
\mathcal{L}_{\rm Dirac},
\end{equation}
with
\begin{equation}
\mathcal{L}_{\rm Dirac}
=
\bar{\psi}
\gamma^I e^\mu_I
\left(
1-\frac{i}{\beta}\gamma^5
\right)
i\nabla_\mu\psi
-
m_\psi \bar{\psi}\psi
+
{\rm h.c.},
\end{equation}
where $\beta$ parametrizes a possible nonminimal coupling
\cite{Mercuri:2006um,Alexander:2014eva,Alexander:2014uaa}.
The spin connection naturally decomposes into a torsion-free piece and
a contorsion contribution,
\begin{equation}
\omega_\mu^{IJ}
=
\omega_\mu^{IJ}[e]
+
C_\mu^{IJ}.
\end{equation}
A crucial feature of Einstein--Cartan gravity is that torsion does not
propagate. The contorsion tensor therefore appears only algebraically
and may be integrated out exactly. Performing this procedure yields an
effective four-fermion interaction
\cite{Freidel:2005sn,PhysRevD.73.044013,
Alexander:2014eva,Alexander:2014uaa},
\begin{equation}
S_{\rm int}
=
-\xi \kappa^2
\int d^4x\, |e|\,
J_5^I J_{5I},
\end{equation}
where
\begin{equation}
J_5^I
=
\bar{\psi}
\gamma^I \gamma^5
\psi
\end{equation}
is the axial current.
The coupling constant is
\begin{equation}
\xi
=
\frac{3}{16}
\frac{\gamma^2}{\gamma^2+1}
\left(
1+\frac{2}{\beta\gamma}
-\frac{1}{\beta^2}
\right),
\end{equation}
whose properties were extensively discussed in
Refs.~\cite{Alexander:2014eva,Freidel:2005sn,deBlas:2013qqa}.
For suitable choices of $\gamma$ and $\beta$, the interaction becomes
attractive and can be substantially stronger than the naive
gravitational coupling. This observation is the key ingredient that
allows gravitational interactions to induce fermion condensation.
To expose the relevant channels, we employ the Fierz identity
\begin{equation}
(\bar{\psi}\gamma_5\gamma^I\psi)
(\bar{\psi}\gamma_{5}\gamma_I\psi)
=
(\bar{\psi}\psi)^2
-
(\bar{\psi}\gamma_5\psi)^2
+
(\bar{\psi}\gamma^I\psi)
(\bar{\psi}\gamma_I\psi),
\label{eq:fierz}
\end{equation}
together with the Pauli--Kofink relation
\cite{PhysRev.140.B1467},
\begin{equation}
(\bar{\psi}\gamma^I\psi)
(\bar{\psi}\gamma_I\psi)
=
(\bar{\psi}\psi)^2
-
(\bar{\psi}\gamma_5\psi)^2.
\end{equation}
The interaction therefore reduces to
\begin{equation}
J_5^I J_{5I}
=
2
\Big[
(\bar{\psi}\psi)^2
-
(\bar{\psi}\gamma_5\psi)^2
\Big].
\end{equation}
The effective fermionic action becomes
\begin{equation}
S_{\rm eff}
=
\int d^4x\, |e|
\left[
\bar{\psi}
i\gamma^\mu\nabla_\mu
\psi
+
\frac{1}{2M_\star^2}
\Big(
(\bar{\psi}\psi)^2
-
(\bar{\psi}\gamma_5\psi)^2
\Big)
\right],
\label{eq:njl_action}
\end{equation}
where
\begin{equation}
\frac{1}{2M_\star^2}
=
-2\xi\kappa^2.
\end{equation}
Equation~(\ref{eq:njl_action}) is structurally identical to the
Nambu--Jona-Lasinio model
\cite{Inagaki:1993ya},
with the important difference that the effective interaction emerges
directly from spacetime geometry rather than being introduced
phenomenologically.
The resulting theory admits two distinct phases.
For weak coupling, the vacuum remains symmetric and no condensate forms.
For sufficiently strong attractive interaction, however, the vacuum
becomes unstable and develops a nonvanishing expectation value
\begin{equation}
\langle \bar{\psi}\psi\rangle
\neq 0.
\end{equation}
The transition between these phases is controlled by a critical
coupling. Introducing a cutoff scale $\Lambda$, the gap equation
implies the existence of a critical value
\begin{equation}
G_c
=
\frac{2\pi^2}{\Lambda^2},
\end{equation}
or equivalently
\begin{equation}
M_{\star,c}^2
=
\frac{\Lambda^2}{2\pi^2},
\end{equation}
which separates the symmetric and condensed phases.
This critical threshold plays a central role throughout the remainder
of the paper. Inflation becomes possible only in the broken phase,
where composite condensate degrees of freedom emerge dynamically.
To describe these collective modes, we introduce bosonic auxiliary
fields through a Hubbard--Stratonovich transformation. The resulting
effective theory contains scalar and pseudoscalar condensate
degrees of freedom whose dynamics govern the inflationary epoch.
In the next section we show how these composite fields arise and how,
when two fermionic sectors are present, they naturally generate a
hybrid-inflation-like structure with a density-driven waterfall
transition.

\section{Composite Condensates and Inflationary Dynamics}
\label{sec:condensates}
\noindent 
The four-fermion interaction derived in the previous section provides
the microscopic origin of the inflationary degrees of freedom.
In the strong-coupling regime, the fermionic vacuum becomes unstable
toward condensation, and the relevant low-energy excitations are no
longer the fundamental fermions themselves but collective composite
modes.

The emergence of these degrees of freedom can be described through a
Hubbard--Stratonovich transformation, which replaces the four-fermion
interaction by auxiliary bosonic fields
\cite{Inagaki:1993ya,Addazi:2017qus}.

For a single fermionic sector, we introduce scalar and pseudoscalar
fields $\Sigma$ and $\Pi$ according to

\begin{equation}
\mathcal{L}_{\rm HS}
=
-\frac{M_\star^2}{2}
\left(
\Sigma^2+\Pi^2
\right)
-
\bar{\psi}
\left(
\Sigma+i\gamma^5\Pi
\right)
\psi .
\end{equation}

The effective fermionic action then becomes

\begin{align}
S_{\Sigma,\Pi}
=
\int d^4x\,|e|
\Big[
&
\bar{\psi}
i\gamma^\mu\nabla_\mu
\psi
\nonumber\\
&
-\bar{\psi}
(\Sigma+i\gamma^5\Pi)
\psi
\nonumber\\
&
-\frac{M_\star^2}{2}
(\Sigma^2+\Pi^2)
\Big].
\end{align}

It is convenient to combine the scalar and pseudoscalar condensates
into the complex field

\begin{equation}
\mathbf{A}
=
\Sigma+i\gamma^5\Pi
=
A\,e^{i\gamma^5\theta},
\end{equation}

with

\begin{equation}
A^2
=
\Sigma^2+\Pi^2.
\end{equation}

Integrating out the fermionic degrees of freedom generates an effective
action for the condensate field,

\begin{equation}
S_{\rm eff}
=
-\int d^4x\,|e|\,
V_{\rm eff}(A)
+
S_{\rm kin}.
\end{equation}

The resulting one-loop effective potential is of
Coleman--Weinberg/Nambu--Jona-Lasinio type
\cite{PhysRevD.29.1584,Inagaki:1993ya,Addazi:2017qus},

\begin{align}
V_{\rm eff}(A)
=
&
V_0
+
\frac{M_\star^2}{2}A^2
\nonumber\\
&
-\frac{1}{4\pi^2}
\Bigg[
A^2\Lambda^2
+
\Lambda^4
\ln
\left(
1+\frac{A^2}{\Lambda^2}
\right)
\nonumber\\
&
\qquad
-
A^4
\ln
\left(
1+\frac{\Lambda^2}{A^2}
\right)
\Bigg]
+
\mathcal{O}(R).
\label{eq:singlepotential}
\end{align}

As shown in Ref.~\cite{Addazi:2017qus}, this potential naturally
supports a slow-roll inflationary phase compatible with current
cosmological observations \cite{Planck:2018jri}.

However, the single-condensate theory suffers from a serous limitation. The field excursion required to sustain inflation is typically
super-Planckian, while the effective theory itself is characterized by
the compositeness scale $\Lambda$.

Furthermore, differently than in our framework, the single-field potential does not naturally provide a
mechanism for fragmentation, reheating, or compact-object formation.
The condensate simply rolls toward its minimum without generating a
waterfall instability.

These observations motivate the extension to multiple condensate
sectors.

\section{Hybrid Inflation from a Fermion Condensate}
\label{sec:hybrid}
\noindent
In this section, we extend the single-field fermion condensate model to a hybrid inflationary framework. To achieve this, we introduce a bipartite fermion sector coupled to gravity. Let the total fermion field be composed of two distinct sets: $\{\Psi_i\}$ with $i=1\dots N$, collectively denoted as $\psi$, and $\{\Psi_j\}$ with $j=1\dots \bar{N}$, denoted as $\chi$. The total action is the sum of the contributions from each constituent fermionic field, i.e. 
\begin{equation}
    \mathcal{S}[\psi]=\sum_{i=1}^{N}\mathcal{S}[{\Psi_i}] \,, \qquad \mathcal{S}[\chi]=\sum_{j=1}^{\bar{N}}\mathcal{S}[{\Psi_j}]\,,
\end{equation}
where each fermion field $\Psi$ is governed by the standard Dirac action. 

Within the Einstein-Cartan-Holst (ECH) framework, substituting the contorsion tensor with its equation of motion generates an effective Lagrangian comprising a kinetic term and a four-fermion interaction, namely 
\begin{align}
    \mathcal{L}_{\rm Dirac}'[\psi, \chi] &= \bar{\psi}\gamma^I e^\mu_I \imath \tilde{\nabla}_\mu\psi+ \bar{\chi}\gamma^I e^\mu_I \imath \tilde{\nabla}_\mu\chi+{\rm h.c.} \,, \\ 
    \mathcal{L}_{\rm int} [\psi, \chi] &= - \xi\kappa^2 J_5^L J_5^M \eta_{LM},
\end{align}
where the axial current is defined as $J^I_5 = \bar{\psi}\gamma_5 \gamma^I \psi + \bar{\psi}\gamma_5 \gamma^I \chi + \bar{\chi}\gamma_5 \gamma^I \psi + \bar{\chi}\gamma_5 \gamma^I \chi$, and $\tilde{\nabla}_\mu$ is the torsion-free, metric-compatible covariant derivative. This bipartition is critical: sequentially integrating out the $\chi$ and $\psi$ sectors yields two distinct composite bound fields. One serves as the slowly rolling inflaton, while the other functions as the auxiliary waterfall field, perfectly mimicking the architecture of hybrid inflation.

\subsection{Bosonization and the Waterfall Mechanism}
\label{subsec:bosonization}

\noindent 
To extract the bound state associated with $\chi$, we follow the methodology of Ref.~\cite{Alexander:2022cow} and introduce an auxiliary field $\alpha$ via a Hubbard--Stratonovich transformation, multiplying the fermion path-integral by the constant term 
\begin{equation}
    Z_\alpha = \int \mathcal{D}\alpha \mathcal{D}\bar{\alpha} \exp\left(-\int d^4x M_\phi^{2}\bar{\alpha}\alpha\right)\,.
\end{equation}
Here, $M_\phi$ represents the bare mass term corresponding to the composite scalar field defined for $\chi$, namely 
\begin{equation}
    \label{eq:boundfield}
    \phi = \alpha - M_\phi^{-2} \bar{\chi}_{R}\chi_{L}.
\end{equation}
Combining $\psi$ and $\chi$ into $\Psi$ for simplicity, the full fermionic Lagrangian becomes
\begin{equation}
    \begin{aligned}
        &\mathcal{L}_{\rm fer} = \bar{\Psi}(i\gamma^\mu\nabla_\mu-m_\psi) \Psi +\frac{1}{2m^2}[(\bar{\Psi}\Psi)^2 + (i\bar{\Psi}\gamma^5\Psi)^2] \\ 
        &\quad + \left[M_\phi^{-2}\bar{\chi}_{L}\chi_{R}\bar{\chi}_{R}\chi_{L} + \left(\phi\bar{\chi}_{R}\chi_{L}+\textrm{h.c.}\right) +  M_\phi^2 |\phi|^2\right]. 
    \end{aligned}
\end{equation}
Integrating out the $\chi$ sector at the one-loop level recasts the effective Lagrangian as \cite{Alexander:2022cow} 
\begin{equation}
    \label{eq:Lefftwof}
    \begin{split}
        \tilde{\mathcal{L}}_{\rm fer} &= \bar{\psi}(i\gamma^\mu\nabla_\mu-m_\psi) \psi + \frac{1}{2m^2}[(\bar{\psi}\psi)^2 + (i\bar{\psi}\gamma^5\psi)^2] \\ 
        &\quad + (\partial_\mu\phi^*)(\partial^\mu\phi) +  y(\phi\bar{\psi}_L\psi_R + \textrm{h.c.}) \\
        &\quad + m_\phi^2|\phi|^2 - \frac{\lambda_\phi}{4}|\phi|^4, 
    \end{split}
\end{equation}
where the induced effective parameters are:
\begin{equation}
    \label{eq:lphi}
    m_\phi^2 = \frac{M_\phi^2 + W_0}{W_2}, \quad y = \frac{W_0}{2W_2^{1/2}m^2}, \quad \lambda_\phi = \frac{V_4}{W_2^2}.
\end{equation}
The parameters $W_0$, $W_2$, and $V_4$ arise from the loop integrals of the fermion propagators \cite{Alexander:2022cow}. For massless fermions, these approximate to $W_0 \approx -\Lambda^2/(2\pi^2)$, $W_2 \approx 1/(2\pi^2)$, and $y \approx -\Lambda^2/(m^2\sqrt{2}\pi)$. Because the bare mass $M_\phi^2$ is not fundamentally constrained, it provides the necessary freedom to tune the effective mass $m_\phi^2$ to support a slow-roll potential. The self-interaction term $\lambda_\phi$, generated by a Yukawa box diagram, is structurally subdominant and can be neglected following dimensional renormalization.

Note that $\phi$ is a complex field. Its real part corresponds to the scalar component, $(\alpha + \bar{\alpha})/2 - \tilde{m}_\phi^{-2} \bar{\chi}\chi/2 \equiv \sigma$, while its imaginary part corresponds to the pseudoscalar component, $(\bar{\alpha} - \alpha)/2 - \tilde{m}_\phi^{-2} (i\bar{\chi}\gamma^5\chi) \equiv \varpi$. 

Applying a subsequent Hubbard--Stratonovich transformation to the primary fermion sector $\psi$, the effective Lagrangian takes the form 
\begin{equation}
    \begin{split}
        \tilde{\mathcal{L}}_\psi &= \bar{\psi}(i\gamma^\mu\nabla_\mu)\psi - \frac{m^2}{2}A^2 - \bar{\psi}(A+y(\sigma+i\gamma^5\varpi))\psi \\ 
        &\quad + (\partial_\mu\phi^*)(\partial^\mu\phi) + m_\phi^2|\phi|^2, 
    \end{split}
\end{equation}
where $A \equiv \Sigma + i\gamma^5\Pi = |A|e^{i\gamma^5\theta_A}$ is the previously defined gap field. For notational convenience, we define $B \equiv \sigma+i\gamma^5\varpi = |B|e^{i\gamma^5\theta_B}$. 

Integrating over the $\psi$ fermion modes generates the Coleman--Weinberg-like effective potential. For clarity, we restrict our analysis to the specific trajectory where $A$ is purely scalar and $B$ is purely pseudoscalar, which is achieved by fixing the chiral phases $\theta_A$ and $\theta_B$. Substituting $|B| = |\phi|$, the resulting effective potential governing this hybrid-like inflationary scenario takes the form 
\begin{equation}
    \label{eq:effpotentialh}
    V(A,B) = V_0 + \frac{m^2}{2}A^2 - m_\phi^2 |B|^2 + \bar{V}(A,B).
\end{equation}
Here, $\bar{V}(A,B)$ represents the standard Nambu--Jona-Lasinio potential \cite{Inagaki:1993ya, Addazi:2017qus}, generalized by the substitution $A \rightarrow A + yB$. The parameter $y$ is an effective coupling constant generated at the one-loop level upon integrating out the second fermionic sector \cite{Alexander:2022cow}. Crucially, the dynamical running of the inflaton field $B$ can trigger a phase transition in the gap field $A$, naturally realizing a waterfall mechanism. This symmetry-breaking trajectory is illustrated in Fig.~\ref{fig:Potential}.

The explicit form of the NJL extension $\bar{V}(A,B)$ is given by
\begin{equation}
    \begin{aligned}
        \bar{V}(A,B) &= - \frac{1}{4\pi^2} \Biggl[ (A^2 + y^2B^2)\Lambda^2 \\
        &\quad + \Lambda^4\ln\left(1+\frac{A^2 + y^2B^2}{\Lambda^2}\right) \\
        &\quad - (A^2+y^2B^2)^2\ln\left(1+\frac{\Lambda^2}{A^2+y^2B^2}\right) \Biggr] \\
        &\quad - \frac{R}{96\pi^2}\Biggl[-(A^2+y^2B^2)\ln\left(1+\frac{\Lambda^2}{A^2+y^2B^2}\right) \\
        &\quad +\frac{\Lambda^2(A^2+y^2B^2)}{\Lambda^2+A^2+y^2B^2} \Biggr]\,.
    \end{aligned}
\end{equation}

\subsection{Parameter Space for Slow-Roll and the Waterfall Transition}

\noindent
To rigorously assess the viability of this potential for inflation, we evaluate the system against the slow-roll and waterfall conditions. We identify $B$ as the slowly rolling inflaton. For a successful waterfall transition, the symmetric state $A=0$ must dynamically destabilize at a critical value of $B$, while the trajectory minimum along $B=0$ must remain globally stable. 

Crucially, the second derivative of the effective potential at the origin, $V''(0)$, exhibits a logarithmic divergence ($\to -\infty$) due to the presence of a sharp Fermi surface. This non-analytic behavior corresponds exactly to the well-known Cooper instability, dictating that the symmetric phase ($A=0$) is unconditionally unstable at zero temperature. Consequently, standard perturbative Taylor expansions around the origin are invalid; dynamical stability must instead be assessed using exact solutions to the gap equations.

The gap equations governing $A$ and $B$ are:
\begin{align}
    \tilde{m}^2 \tilde{A} &= \frac{\tilde{A}}{\pi^2} \left[ 1 - (\tilde{A}^2 + y^2 \tilde{B}^2) \ln\left(1 + \frac{1}{\tilde{A}^2 + y^2 \tilde{B}^2}\right) \right] \notag \\ 
    &\quad - \frac{\tilde{R}\tilde{A}}{48\pi^2} \Biggl[ -\ln\left(1 + \frac{1}{\tilde{A}^2 + y^2 \tilde{B}^2}\right) \notag \\
    &\quad + \frac{\tilde{A}^2 + y^2 \tilde{B}^2 + 2}{(\tilde{A}^2 + y^2 \tilde{B}^2 + 1)^2} \Biggr]\,, \label{eq:gapA} \\ 
    2\tilde{m}_\phi^2 \tilde{B} &= - \frac{y^2 \tilde{B}}{\pi^2} \left[ 1 - (\tilde{A}^2 + y^2 \tilde{B}^2) \ln\left(1 + \frac{1}{\tilde{A}^2 + y^2 \tilde{B}^2}\right) \right] \notag \\ 
    &\quad - \frac{\tilde{R}y^2\tilde{B}}{48\pi^2} \Biggl[ -\ln\left(1 + \frac{1}{\tilde{A}^2 + y^2 \tilde{B}^2}\right) \notag \\
    &\quad + \frac{\tilde{A}^2 + y^2 \tilde{B}^2 + 2}{(\tilde{A}^2 + y^2 \tilde{B}^2 + 1)^2} \Biggr], \label{eq:gapB}
\end{align}
where we have normalized the fields and parameters by the cutoff scale (e.g., $\tilde{A}\equiv A/\Lambda$, $\tilde{B} \equiv B/\Lambda$). 

During inflation, the spacetime curvature scales as $R\propto H^2$. The Friedmann equation yields, namely 
\begin{equation}
    H^2 = \frac{1}{3M_{\rm pl}^2}V(A,B) \approx \frac{\Lambda^4}{3M_{\rm pl}^2} \ll \Lambda^2.
\end{equation}
This demonstrates that the Hubble scale $H$ is deeply suppressed relative to the energy cutoff $\Lambda$. We can therefore safely neglect the curvature contributions in the gap equations. Defining $\tilde{m}^2 = (1-\delta)/\pi^2$, the solutions to the simplified gap equations fall along $\tilde{B} = 0$ in conjunction with either $\tilde{A} = 0$ or the following transcendental constraints:
\begin{align}
    \frac{\delta}{\tilde{A}^2+y^2\tilde{B}^2} &\approx \ln\left(1+\frac{1}{\tilde{A}^2+y^2\tilde{B}^2}\right), \label{eq:BERT1} \\ 
    \frac{1+4(1-\delta)^2\tilde{m}_\phi^2}{\tilde{A}^2 + y^2\tilde{B}^2} &\approx \ln\left(1+\frac{1}{\tilde{A}^2 + y^2\tilde{B}^2}\right). \label{eq:BERT2}
\end{align}
To ensure that the waterfall transition of the field $A$ is physically executed, the global minimum of the potential must map to $\tilde{B} = 0$ alongside a dynamically generated $A\neq 0$. This requires Eq.~\eqref{eq:BERT1} to be satisfied while Eq.~\eqref{eq:BERT2} is excluded, strictly bounding the parameter space to $0 < \delta < 1$ and $\tilde{m}_\phi^2 \le -\frac{1}{4(1-\delta)}$. The visualization in Fig.~\ref{fig:Potential} adopts $\delta = 0.5$ and $\tilde{m}_\phi^2 = -1$ to clearly trace this phase transition.

Finally, we constrain the parameter space against the standard slow-roll conditions:
\begin{equation}
    \epsilon = \frac{M_{\rm pl}^2}{2}\left(\frac{V'(B)}{V(B)}\right)^2\ll 1, \quad \eta = M_{\rm pl}^2\frac{V''(B)}{V(B)}\ll 1.
\end{equation}
Direct calculation demonstrates that the $\epsilon \ll 1$ condition is robustly satisfied for $B \ll 10^{-3}\Lambda$, independent of the specific values of $\delta$ and $m_\phi$. Satisfying the $\eta$ condition requires tighter constraints. The exact second derivative of the potential with respect to the inflaton $B$ is
\begin{equation}
    \label{eq:2ndd}
    \begin{split}
        \frac{\partial^2 V}{\partial B^2} &= -2m_\phi^2 - \frac{y^2}{\pi^2} \Biggl[ \Lambda^2 + \frac{2y^2B^2\Lambda^2}{A^2+y^2B^2+\Lambda^2} \\
        &\quad - (A^2+3y^2B^2)\ln\left(1+\frac{\Lambda^2}{A^2+y^2B^2}\right) \Biggr] \\ 
        &\quad - \frac{R y^2}{48\pi^2} \Biggl[ -\ln\left(1+\frac{\Lambda^2}{A^2+y^2B^2}\right) \\
        &\quad + \frac{\Lambda^2(A^2+y^2B^2+2\Lambda^2)}{(A^2+y^2B^2+\Lambda^2)^2} \\ 
        &\quad + \frac{2y^2B^2\Lambda^2}{(A^2+y^2B^2)(A^2+y^2B^2+\Lambda^2)} \\
        &\quad - \frac{2y^2B^2\Lambda^2(A^2+y^2B^2+3\Lambda^2)}{(A^2+y^2B^2+\Lambda^2)^3} \Biggr]. 
    \end{split}
\end{equation}
To simultaneously align the slow-roll and waterfall requirements, two distinct kinematic scenarios exist: either (1) the waterfall occurs \textit{after} the slow-roll phase (meaning $A=0$ during inflation); or (2) the waterfall occurs \textit{before} the slow-roll phase, allowing inflation to proceed on the broken symmetry branch. In both regimes, the effective mass $m_\phi^2$ can be explicitly tuned to cancel the dominant constant terms in the second derivative. This leaves a residual restoring force proportional to $y^2B^2$, preserving $\eta \ll 1$ strictly within the domain $B \lesssim 10^{-4}\Lambda$ ---  the scenario 1 additionally requires extreme parameter suppression, i.e. $\delta \sim \mathcal{O}(10^{-10})$. 

However, achieving a graceful exit from this inflationary phase requires additional dynamics. In the subsequent section, we introduce an axial chemical potential to account for continuous particle production, establishing a density-driven mechanism to terminate inflation and trigger instant preheating.

\section{Particle Production, Phase Transition, and the End of Inflation}
\label{sec:end_inflation}

\noindent
The axial $U(1)_A$ symmetry, defined by the transformation $\psi\rightarrow e^{i\gamma^5\theta}\psi$, is explicitly broken in the effective action by the presence of the macroscopic condensate term, $\bar{\psi}(\Sigma+i\gamma^5\Pi)\psi$. Consequently, the Noether current associated with $U(1)_A$ is not conserved, permitting continuous net fermion number production. To quantify this thermodynamic effect during the inflationary expansion, we introduce a parity-odd axial chemical potential, $\mu$, into the Lagrangian, i.e. 
\begin{equation}
    \mathcal{L}_\mu = -\mu\bar{\psi}\gamma^0\gamma^5\psi\,.
\end{equation}
This term dynamically biases the vacuum, triggering the continuous creation of fermions during inflation \cite{Adshead:2015kza,Chen:2018xck,Adshead:2018oaa,Hook:2019zxa,Sou:2021juh,Chen:2023txq}. Since the chemical potential inherently breaks Lorentz symmetry, the standard derivation of the effective potential via Riemann normal coordinates and Green's functions is no longer applicable. Instead, we derive the finite-density effective potential using the alternative partition-function approach detailed in Appendix~\ref{app:finite_density}. 

The condensate must satisfy the stationary gap equation, $\partial V(A)/\partial A =0$. Expressed in terms of the normalized fields $\tilde{A}$, $\tilde{\mu}$, and $\tilde{H}$, this consistency condition takes the form
\begin{equation}
    \label{eq:gapeq}
    \begin{split}
        \frac{1}{\lambda} &= \frac{1}{\pi^3} \int_0^\Lambda |\vec{q}|^2 d|\vec{q}| \Biggl[ \frac{1}{E_+} \arctan\left(\frac{W_q}{E_+}\right) \\
        &\quad + \frac{1}{E_-} \arctan\left(\frac{W_q}{E_-}\right) \Biggr],
    \end{split}
\end{equation}
where $W_q \equiv \sqrt{\Lambda^2 - |\vec{q}|^2}$ and the modified dispersion relations are $E_\pm = \sqrt{(|\vec{q}| \pm \mu)^2 + A^2}$. 

Defining $U(A) = V(A) - m^2A^2/2$ as the fermionic determinant contribution to the effective potential, we can extract the macroscopic observables by differentiating the partition function with respect to the relevant source variables. This yields the fundamental thermodynamic relations
\begin{align}
    \langle n_A\rangle &= \langle \bar{\psi}\gamma^0\gamma^5\psi\rangle = \left\langle\frac{\delta U}{\delta \mu}\right\rangle\,,\label{eq:prim}\\
    \langle i\bar{\psi}\gamma^5\psi\rangle &= \left\langle\frac{\delta U(A)}{\delta \Pi}\right\rangle = \left\langle\frac{\delta U(A)}{\delta A}\frac{\delta A}{\delta \Pi}\right\rangle\,.\label{eq:psconeq}
\end{align}
Evaluating Eq.~\eqref{eq:prim} provides the governing equation of motion for the axial chemical potential, capturing the phase-space density of the states, namely
\begin{equation}
    \label{eq:Chemicaleq}
    \begin{split}
        \langle n_A\rangle &= \frac{1}{\pi^3} \int_0^\Lambda |\vec{q}|^2 d|\vec{q}| \Biggl[ \frac{|\vec{q}| + \mu}{E_+} \arctan\left(\frac{W_q}{E_+}\right) \\
        &\quad - \frac{|\vec{q}| - \mu}{E_-} \arctan\left(\frac{W_q}{E_-}\right) \Biggr] \,.
    \end{split}
\end{equation}

To resolve the dynamics of Eq.~\eqref{eq:psconeq}, we couple it to the covariant divergence of the $U(1)_A$ Noether current, $J^{\mu}_5=\bar{\psi}\gamma^\mu\gamma_5\psi$, namely 
\begin{equation}
    \label{eq:axialcurrenteq}
    \nabla_\mu \langle J^{\mu}_5\rangle = 2A_0\langle i \bar{\psi}\gamma_5\psi\rangle\,.
\end{equation}
Combining Eq.~\eqref{eq:psconeq} and Eq.~\eqref{eq:axialcurrenteq} recovers the macroscopic chiral anomaly equation embedded within a Friedmann--Lema\^{i}tre--Robertson--Walker (FLRW) spacetime:
\begin{equation}
    \label{eq:anomalyeq}
    \partial_\tau\langle n_A\rangle + 3\mathcal{H}\langle n_A\rangle = \frac{2A_0^2a}{\lambda}\,,
\end{equation}
where $\tau = \mathcal{H}^{-1} = 1/(aH)$ represents conformal time, a relation valid during the subsequent radiation-dominated epoch. Assuming $a \propto \tau$ during this phase, the general solution to Eq.~\eqref{eq:anomalyeq} is
\begin{equation}
    \begin{split}
        \langle n_A\rangle &= \tau^{-3}\left(\text{const.} + 2\int \frac{A_0^2\tau^5}{\lambda} d\tau\right) \\
        &= \tau^{-3}\left(\text{const.} + \frac{2}{5}\frac{A_0^2\tau^5}{\lambda}\right)\,.
    \end{split}
\end{equation}
The initial homogeneous term decays rapidly as $a^{-3}$ and can be safely neglected, whereas the forced term grows proportionally with $\tau^2$. Consequently, the continuous pumping from the anomaly ensures that the macroscopic axial number density $\langle n_A\rangle$ strictly increases over time.

Rather than attempting to solve Eqs.~\eqref{eq:gapeq} and \eqref{eq:Chemicaleq} simultaneously for an arbitrary target density --- a procedure that obscures the global topology of the solution space --- we systematically map the accessible phase space over a high-resolution 2D grid spanning the dimensionless coordinates $(\tilde{A}, \tilde{\mu})$. 

We define a gap equation residual, $\mathcal{R}_{\rm gap}(\tilde{A}, \tilde{\mu})$, which quantifies the deviation from the true vacuum state at any arbitrary point in the parameter space, i.e. 
\begin{equation}
    \begin{split}
        \mathcal{R}_{\rm gap}(\tilde{A}, \tilde{\mu}) &= \frac{1}{\pi^3} \int_0^\Lambda |\vec{q}|^2 d|\vec{q}| \Biggl[ \frac{1}{E_+} \arctan\left(\frac{W_q}{E_+}\right) \\
        &\quad + \frac{1}{E_-} \arctan\left(\frac{W_q}{E_-}\right) \Biggr] - \frac{1}{\tilde{\lambda}}\,.
    \end{split}
\end{equation}
The physical vacuum manifold is defined strictly by the locus of points where $\mathcal{R}_{\rm gap}(\tilde{A}, \tilde{\mu}) = 0$. 

The numerical integration over the momentum variable $|\vec{q}|$ is performed concurrently across the 2D grid using a vectorized Simpson's rule. This evaluates both the gap residual $\mathcal{R}_{\rm gap}$ and the density field $\tilde{n}_A(\tilde{A}, \tilde{\mu})$ simultaneously, bypassing the computational overhead of sequential nested loops. 

By extracting the zero-contour of $\mathcal{R}_{\rm gap}$, we isolate the physically permitted evolutionary pathways of the Universe. Superimposing the contours of constant $\tilde{n}_A$ onto this manifold, as illustrated in Fig.~\ref{fig:condensateEOM}, reveals the existence of a strict topological threshold. As particle production drives the axial density higher, the system traverses the red vacuum manifold until it reaches a tangent maximum, $\tilde{n}_{A,\text{crit}}$. 

As clearly shown in Fig.~\ref{fig:condensateEOM}, for normalized densities $\langle n_A\rangle /\Lambda^3 \gtrsim 0.035$, the constant density contours permanently decouple from the real-valued vacuum manifold; no intersections exist. Beyond this critical limit, the energetic cost of sustaining the coherent condensate exceeds the available phase space. This decoupling signals the abrupt restoration of chiral symmetry ($\tilde{A} \to 0$) and marks a natural, graceful exit from the condensate-driven inflationary phase.

\subsection{Physical Picture: The Dynamical Waterfall}
\label{sec:dynamical_exit}

\noindent
Unlike standard hybrid inflation, where the exit is triggered geometrically when the inflaton crosses a critical spatial threshold in field space, the exit mechanism in this NJL framework is intrinsically thermodynamic. It is governed by the temporal accumulation of the axial particle number density. During inflation, continuous particle production drives up the chiral chemical potential $\mu(t)$, which acts as an effective Fermi energy, gradually filling a dense Fermi sea of real fermions.

As this Fermi sea deepens, the real fermions exert a powerful degeneracy pressure, invoking strict Pauli blocking. This statistical pressure physically prevents the virtual fermions in the Dirac sea from pairing into the coherent state required to sustain the macroscopic chiral condensate, $\langle \bar{\psi}\psi \rangle$. Once the chemical potential approaches the cutoff scale ($\mu(t) \sim \Lambda$), the gap equation mathematically loses its real-valued solution. Thermodynamically, the free energy cost of maintaining the symmetry-broken state surpasses that of a free Fermi gas. The condensate therefore instantaneously ``melts'' ($A \to 0$), executing a density-driven phase transition. We formally designate this synthesis of field-space instability and thermodynamic condensate melting as the \textit{dynamical waterfall mechanism}.

The abrupt destruction of the false vacuum induces a violent tachyonic instability, liberating the stored inflationary vacuum energy. The precise microscopic dynamics governing how this energy is instantaneously transferred back into the fermionic plasma to successfully reheat the Universe is detailed in the following section.

\section{Instant Preheating and Energy Transfer}
\label{sec:preheating}

\noindent 
The density-driven waterfall discussed in Sec.~\ref{sec:dynamical_exit} terminates inflation through the thermodynamically favored evaporation of the fermion condensate. A critical requirement for any viable inflationary model is the efficient transfer of this stored vacuum energy into relativistic degrees of freedom to successfully reheat the Universe.

In conventional models driven by elementary scalar fields, reheating typically proceeds through perturbative inflaton decay or non-perturbative parametric resonance \cite{Traschen:1990sw,Kofman:1994rk,Kofman:1997yn}. In the present framework, the dynamical situation is fundamentally different. The field responsible for the vacuum energy is not elementary, but rather a collective macroscopic excitation of the fermionic medium itself. Consequently, the degrees of freedom into which the condensate decays are already present in the microscopic theory. This economy of ingredients ensures that the density-driven waterfall naturally provides a microscopic realization of instant preheating.

\subsection{Condensate Collapse and Effective Decay}
\noindent 
At the transition point ($\mu \sim \Lambda$), the finite-density gap equation ceases to admit a broken-phase solution. The effective potential undergoes a rapid topological rearrangement, abruptly restoring chiral symmetry ($A \to 0$). Prior to this transition, the energy density of the universe is dominated by the vacuum energy of the condensate, $\rho_{\rm vac} \simeq V_0 \simeq \Lambda^4$. Once the condensate branch vanishes, this energy can no longer be supported by the symmetric vacuum and must be violently redistributed among the microscopic fermionic degrees of freedom.

Since the gap field $A$ originates from the bosonization of the fermion bilinear, the effective action inherently contains a direct Yukawa interaction, i.e. 
\begin{equation}
    \mathcal{L}_{\rm Yuk} = - y_{\rm eff} A \bar{\psi}\psi.
\end{equation}
Unlike phenomenological scalar inflation models where decay couplings are introduced by hand, this interaction is an unavoidable consequence of the condensate construction. Being the condensate generated by strong four-fermion interactions, the effective Yukawa coupling is naturally of order unity, $y_{\rm eff} = \mathcal{O}(1)$.

The decay rate of the massive gap fluctuations into fermion pairs is approximately
\begin{equation}
    \Gamma_{A\rightarrow\bar{\psi}\psi} \simeq \frac{y_{\rm eff}^2}{8\pi} m_A^{\text{eff}},
\end{equation}
where $m_A^{\text{eff}}$ is the physical effective mass of the scalar fluctuations. Crucially, since $A$ is not a fundamental field but a composite resonance governed by NJL dynamics, its effective mass is determined by the curvature of the full quantum effective potential, $m_A^2 = \partial^2 V_{\rm eff} / \partial A^2$. Following standard NJL phenomenology, the mass of this composite scalar is strictly tied to the dynamically generated gap scale, $m_A^{\text{eff}} \approx 2 A_{\rm vac}$. Requiring the gap equation $A_{\rm vac} \sim \Lambda$, the effective mass naturally scales as $m_A^{\text{eff}} \sim \Lambda$. 

Combined with an $\mathcal{O}(1)$ effective Yukawa coupling, the decay rate robustly scales as $\Gamma_A \sim 10^{-1}\Lambda$.

To determine the efficiency of this energy transfer, we compare the decay rate to the Hubble expansion rate at the end of inflation, $H_{\rm end} \simeq \Lambda^2 / (\sqrt{3}M_{\rm pl})$. For a characteristic inflationary energy scale of $\Lambda \sim 10^{-3} M_{\rm pl}$ (i.e., the Grand Unified Theory scale), the Hubble parameter is heavily suppressed by the Planck mass, yielding $H_{\rm end} \sim 10^{-6} M_{\rm pl}$. 

In contrast, the decay rate is $\Gamma_A \sim 10^{-4} M_{\rm pl}$. The efficiency ratio is therefore
\begin{equation}
    \frac{\Gamma_A}{H_{\rm end}} \sim 10^2 \gg 1.
\end{equation}
Being the decay process orders of magnitude faster than the cosmological expansion rate, the phase transition operates as an effectively instantaneous energy-transfer mechanism.

\subsection{Thermalization and Reheating Temperature}
\noindent 
Following the instant decay of the condensate, the resulting dense fermionic plasma must thermalize to establish a standard radiation-dominated era. The produced fermions continue to interact via the underlying torsion-induced four-fermion operators that initially drove the condensation. The effective interaction cross-section scales as $\sigma \sim s / M_\star^4$, where $s$ is the Mandelstam invariant.

The interaction rate is given by $\Gamma_{\rm scat} \sim n_\psi \sigma$. Immediately after the waterfall transition, the number density is governed by the energy scale of the transferred vacuum, $n_\psi \sim \Lambda^3$. This yields a scattering rate $\Gamma_{\rm scat} \gg H_{\rm end}$, guaranteeing that the fermionic plasma rapidly achieves local thermal equilibrium. The thermalization process is therefore elegantly driven by the exact same microscopic interactions responsible for the initial inflationary condensation.

Assuming an efficient conversion of the vacuum energy into a thermal bath, the reheating temperature $T_{\rm rh}$ is derived from the standard radiation density relation, 
\begin{equation}
    \rho_{\rm vac} = \frac{\pi^2}{30} g_\ast T_{\rm rh}^4,
\end{equation}
where $g_\ast$ denotes the effective number of relativistic degrees of freedom. Substituting $\rho_{\rm vac} \simeq \Lambda^4$, we find
\begin{equation}
    T_{\rm rh} \simeq \left( \frac{30}{\pi^2 g_\ast} \right)^{1/4} \Lambda.
\end{equation}
For a typical Standard Model (or mild extension) value of $g_\ast \sim 10^2$, this yields $T_{\rm rh} \simeq 0.4\,\Lambda$. Thus, for the characteristic inflationary scale of $\Lambda \sim 10^{15}-10^{16}\ {\rm GeV}$, the reheating temperature naturally lies within the range of $10^{15}\ {\rm GeV}$.

In summary, the density-driven waterfall acts as a finite-density phase transition within a strongly interacting fermionic medium. The temporal accumulation of axial charge drives the system to a critical density threshold, forcing the symmetry-broken state to evaporate. The vacuum energy is instantly transferred into a thermal fermion plasma without the need for an independent, fine-tuned reheating sector. Furthermore, the rapid collapse of this condensate generates large spatial inhomogeneities. As we will explore in the following section, these fluctuations subsequently seed the fragmentation of the field, leading to the formation of non-topological solitons (Q-balls) and providing a direct mechanism for primordial black hole production.

\section{Fragmentation into Q-Balls}
\label{sec:qballs}

\noindent 
The density-driven waterfall terminates inflation through the thermodynamic evaporation of the homogeneous condensate. However, this phase transition does not merely convert vacuum energy into a uniform thermal bath. As the macroscopic field is rapidly displaced from its symmetric false vacuum, it generates massive spatial inhomogeneities. The post-inflationary configuration becomes violently unstable against fragmentation. In this section, we demonstrate that these tachyonic instabilities naturally shatter the condensate into stable, non-topological solitons --- specifically, Q-balls \cite{Coleman:1985ki,kusenko1997small,Kusenko:1997vp,Kusenko:1997si,Dvali:1997qv,Kusenko:2005du}.

\subsection{Tachyonic Instability and the Complex Condensate Field}
\noindent 
The composite condensate sector, generated via the Hubbard-Stratonovich transformation, inherently contains both scalar ($\Sigma$) and pseudoscalar ($\Pi$) degrees of freedom. It is therefore highly convenient to parameterize this system using a complex scalar field,
\begin{equation}
    \label{eq:complexfield}
    \varphi = \rho\,e^{i\theta},
\end{equation}
where the radial mode $\rho$ dictates the amplitude of the condensate and the phase $\theta$ corresponds to the chiral angle. The effective low-energy Lagrangian takes the standard form $\mathcal{L}_\varphi = |\partial_\mu\varphi|^2 - V(|\varphi|^2)$, where $V(|\varphi|^2)$ is the condensate effective potential derived in Sec.~\ref{sec:condensates}.

This complex field possesses an approximate global $U(1)_A$ symmetry, $\varphi \rightarrow e^{i\alpha}\varphi$, associated with the chiral axial charge we previously analyzed. The corresponding conserved Noether charge is given by 
\begin{equation}
    Q = \frac{1}{2i} \int d^3x \left( \varphi^\ast\dot{\varphi} - \varphi\dot{\varphi}^\ast \right).
\end{equation}
This macroscopic charge provides the crucial stabilizing ingredient required for Q-ball formation.

Following the onset of the waterfall transition, the homogeneous condensate is destabilized and rolls aggressively toward the origin. During this trajectory, the field traverses a region where the effective potential exhibits negative curvature, $V''(\rho) < 0$. This tachyonic regime is the precise thermodynamic condition required to trigger the exponential growth of long-wavelength spatial fluctuations \cite{Kusenko:1997si,Kasuya:2000wx,Kasuya:2001hg,Cotner:2019ykd}. 

Decomposing the field into a homogeneous background and Fourier-space perturbations, $\varphi = \varphi_0 + \delta\varphi$, the fluctuations evolve as $\delta\varphi_k \propto e^{\alpha_k t}$. The instability condition ($\alpha_k > 0$) delineates a specific tachyonic amplification band bounded by
\begin{equation}
    \label{eq:instabilityband}
    0 < \frac{k^2}{a^2} < \omega^2 - V''(\rho),
\end{equation}
where $\omega = \dot{\theta}$. As the modes within this momentum band undergo rapid exponential growth, they quickly become nonlinear. The homogeneous condensate shatters, transferring its localized charge and energy into spatially bound lumps.
\begin{figure}[t]
    \centering
    \includegraphics[width=0.45\textwidth]{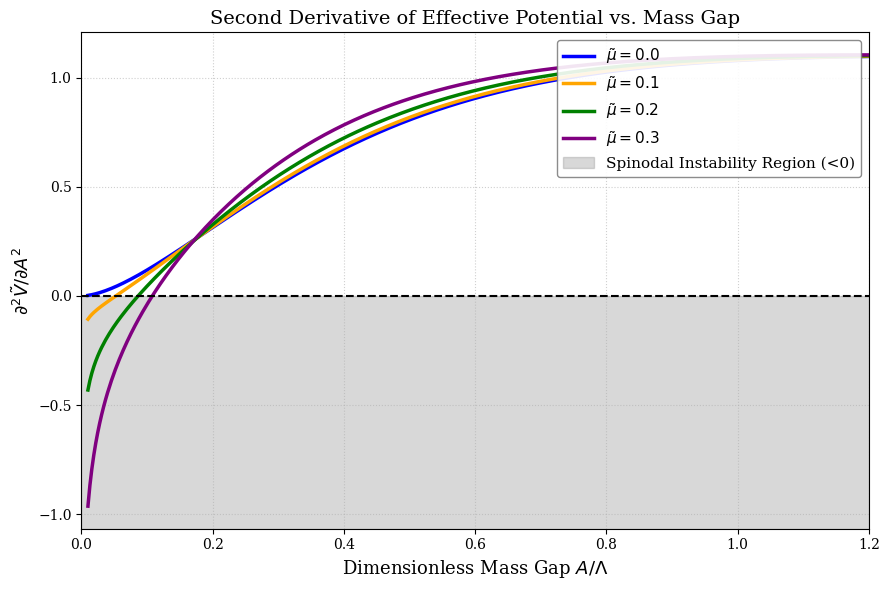}
    \caption{We plot the values of the second derivative of the normalized effective potential, selecting different values of the chemical potential. The negative values of the second derivative are marked in gray, below the dashed horizontal line. We observe that, the potential encompasses a part with negative values, and then becomes positive. This implies, the effective potential is not stable when the gap field is small. The instability then grows, being stabilized only when the gap field becomes larger.}
    \label{fig:2nddV}
\end{figure}

\subsection{Q-Ball Profiling and Scaling Relations}
\noindent 
The nonlinear endpoint of this fragmentation process is a robust population of Q-balls. Following the standard construction \cite{Coleman:1985ki,Dvali:1997qv,Kusenko:1997si}, we isolate these configurations using the stationary ansatz $\varphi = \rho(r)e^{i\omega t}$. The conserved charge evaluates to $Q = \omega \int d^3x\,\rho^2$, and the functional energy of the configuration is $E_\omega = \omega Q + S_\omega$, where
\begin{equation}
    S_\omega = \int d^3x \left[ \frac{1}{2}(\nabla\rho)^2 + V(\rho) - \frac{1}{2}\omega^2\rho^2 \right].
\end{equation}
Minimizing this action yields the radial profile equation
\begin{equation}
    \label{eq:qprofile}
    \rho'' + \frac{2}{r}\rho' + \omega^2\rho - \frac{dV}{d\rho} = 0.
\end{equation}
Because the fermionic determinant generates a Coleman-Weinberg-like potential that flattens out at large field values ($V \rightarrow V_0 \sim \Lambda^4$), the resulting solitons belong to the well-studied ``flat-potential'' class of Q-balls \cite{Dvali:1997qv,Kusenko:2005du,Cotner:2019ykd}.

Energy minimization for flat-potential Q-balls enforces the dynamical scaling $\omega \propto \Lambda Q^{-1/4}$, which immediately dictates the macroscopic properties of the solitons:
\begin{equation}
    \label{eq:scaling}
    E_Q \propto \Lambda Q^{3/4}, \qquad R_Q \propto \frac{Q^{1/4}}{\Lambda}.
\end{equation}
Crucially, the energy-per-charge ratio scales as $E_Q/Q \propto \Lambda Q^{-1/4}$. Because the Q-balls couple directly to the underlying microscopic fermions via the fundamental NJL interaction, they can theoretically decay into fermionic pairs \cite{Cohen:1986ct,Kawasaki:2012gk,Enqvist:1998xd,Kusenko:2005du,Kasuya:2014ofa,Cotner:2016dhw}. However, this decay is only kinematically permitted when the mass of the final states is less than the energy-per-charge ratio, $m_\psi < E_Q/Q$. 

Since $E_Q/Q$ monotonically decreases with $Q$, sufficiently large Q-balls are completely shielded against decay by strict kinematic blocking. Consequently, a substantial fraction of the fragmented Q-ball population forms extremely long-lived compact objects, setting the perfect initial conditions for dark matter relic generation.

We emphasize that the formation of Q-balls is not an ad hoc addition to the model, but a deterministic, mathematical consequence of the density-driven waterfall. The tachyonic instability band organically fragments the symmetry-breaking vacuum, depositing the inflationary energy into a resilient population of Q-balls, which ultimately seed the primordial black holes discussed in the main text.

\section{Primordial Black Holes}
\label{sec:pbh}

\noindent 
One of the most profound phenomenological consequences of the density-driven waterfall mechanism is the deterministic production of solitonic relics. The sequence of early-universe events is strictly continuous: the thermodynamic melting of the condensate triggers tachyonic fragmentation, which deposits the inflationary vacuum energy directly into massive, charge-stabilized Q-balls. 

As these macroscopic lumps come to dominate the local energy density, their subsequent stochastic clustering can trigger gravitational collapse, seeding a population of Primordial Black Holes (PBHs). This provides a rare, direct theoretical bridge linking the microscopic chiral dynamics of ECH gravity to macroscopic, observable astrophysical objects.

\subsection{Stochastic Clustering and Collapse Criteria}
\noindent 
While many inflationary models rely on fine-tuned enhancements in the primordial power spectrum to generate PBHs \cite{Garcia-Bellido:1996mdl,Inomata:2016rbd,Inomata:2017okj,Georg:2017mqk}, the present framework utilizes the intrinsic lumpiness of scalar-field fragmentation \cite{Cotner:2016dhw,Cotner:2016cvr,Cotner:2017tir,Cotner:2019ykd}. Because the spatial phase of the tachyonic instability is inherently stochastic, the exact number of Q-balls fragmenting into a given Hubble volume fluctuates \cite{Kusenko:1997si,Kasuya:2000wx,Multamaki:2002hv}.

We model the large-scale Q-ball distribution using a spatial Poisson process:
\begin{equation}
    \label{eq:poisson}
    P(N|V) = \frac{(\bar n V)^N}{N!} e^{-\bar n V},
\end{equation}
where $\bar n$ is the average Q-ball number density. The total mass contained within a local cluster is simply $M = N\,M_Q$, where the unit mass $M_Q = E_Q \propto \Lambda Q^{3/4}$. These localized statistical fluctuations in $N$ generate extreme energy density spikes, forming massive overdense clusters.

Gravitational collapse into a PBH occurs strictly if a cluster's physical size is confined within its own Schwarzschild radius, $R_{\rm BH} = 2M/M_{\rm pl}^2$ \cite{Zeldovich:1967lct,hawking1971gravitationally,carr1974black,1975ApJ...201....1C,Khlopov:2008qy,Carr:2016drx}. We define a probability filtering function, $B(M,V)$, which measures the exact fraction of clusters of mass $M$ and volume $V$ that satisfy this geometric collapse threshold \cite{Kodama:1986ud,Cotner:2019ykd}. 

The average macroscopic energy density bound in PBHs is then derived by integrating over the statistical distribution, i.e.
\begin{equation}
    \langle\rho_{\rm PBH}\rangle = \int \frac{dV}{V^2} \int dM\, P(M|V)\, B(M,V)\, M.
\end{equation}
This integral completely dictates the terminal PBH abundance, transferring the stochastic nature of the quantum field fragmentation directly into the astrophysics of dark matter.

\subsection{The PBH Mass Function and Dark Matter}
\noindent 
The primary observable for any dark matter candidate is the differential mass function, $df_{\rm PBH}/d\ln M$, which quantifies the fraction of the dark matter energy density bound in PBHs of a specific mass. In our framework, this mass function is not governed by generic inflationary slow-roll parameters, but rather inherits its precise shape from three microscopic inputs: the charge distribution generated during chiral fragmentation, the Q-ball scaling relations --- see e.g. Eq.~\eqref{eq:scaling} --- and the geometric probability of collapse.

The fundamental energy scale of the NJL interaction introduces a strict ultraviolet limit, $\Lambda \sim 10^{-3}M_{\rm pl} \sim 10^{15}-10^{16}\ {\rm GeV}$. This high energy scale, passed through the Q-ball scaling laws, heavily influences the characteristic mass of the resulting PBHs. However, due to the stochastic nature of the Poissonian clustering, the model generically predicts an extended mass spectrum rather than a monochromatic spike. This is phenomenologically vital, as current microlensing and evaporation constraints on PBHs are strongly mass-dependent \cite{Carr:2016drx,Sasaki:2016jop}. 

By varying the fragmentation efficiency and the mean Q-ball charge distribution, the resulting total fractional abundance, $f_{\rm PBH} = \Omega_{\rm PBH} / \Omega_{\rm DM}$, interpolates smoothly. The model gracefully accommodates scenarios where PBHs act as a fractional dark matter constituent ($f_{\rm PBH} \ll 1$) to scenarios where they account for the entire dark matter budget ($f_{\rm PBH} \simeq 1$).

Ultimately, this mechanism uniquely interlocks multiple sectors of physics. The density-driven waterfall simultaneously melts the inflaton condensate, generates massive Q-ball relics, and leaves behind an extended PBH mass function. Furthermore, because the underlying ECH theory is intrinsically chiral, the same fundamental torsion interactions that drive the PBH formation also excite the gravitational sector. This naturally points toward the generation of parity-violating gravitational waves \cite{Alexander:2022cow}. We devote the next section to deriving these gravitational-wave signatures, establishing a secondary, complementary observational probe to definitively test the fermionic origins of inflation.

\section{Parity-Violating Gravitational-Wave Signatures}
\label{sec:gws}

\noindent 
The density-driven waterfall mechanism generates a rich spectrum of observable consequences. While the previous section demonstrated that post-inflationary condensate fragmentation naturally seeds primordial black holes (PBHs), compact-object production is not the sole signature of this framework. A defining feature of torsion-induced fermion condensation is its intrinsically chiral nature. The same microscopic fermionic interactions that drive the inflationary expansion naturally induce parity-violating couplings within the gravitational sector. Consequently, the framework predicts distinctive, observable signatures in both primordial and stochastic gravitational-wave (GW) backgrounds. These signatures provide a direct, multi-messenger probe of the microscopic origin of inflation.

\subsection{Emergent Dynamical Chern-Simons Gravity and Chiral Tensors}
\noindent 
The fundamental connection between fermionic condensates and parity-violating gravity arises directly from the Einstein-Cartan-Holst (ECH) action \cite{Alexander:2022cow}. Upon integrating out the fermionic degrees of freedom, the chiral anomaly and the background torsion inevitably generate a dynamical Chern-Simons (dCS) contribution to the low-energy effective action:
\begin{equation}
    S_{\rm dCS} = \int d^4x \sqrt{-g} \, \frac{\vartheta}{4} \,{}^\ast RR,
    \label{eq:dcs}
\end{equation}
where ${}^\ast RR = \frac{1}{2} \epsilon^{\mu\nu\rho\sigma} R_{\mu\nu\alpha\beta} R_{\rho\sigma}{}^{\alpha\beta}$ is the Pontryagin density, and $\vartheta$ denotes an effective pseudoscalar degree of freedom. In the context of the density-driven waterfall, the role of $\vartheta$ is dynamically played by the chiral phase of the condensate sector --- the angle $\theta$ of the complex field $\varphi$ introduced in Eq.~\eqref{eq:complexfield}. 

The emergence of Eq.~\eqref{eq:dcs} explicitly breaks parity as an exact symmetry of the gravitational sector. As a result, the tensor perturbations of the metric, decomposed into left- and right-handed circular polarization states ($h_{ij} = h_R e_{ij}^{R} + h_L e_{ij}^{L}$), experience asymmetric evolution. In standard parity-conserving theories, the power spectra for these states are identical --- namely, $P_h^R(k) = P_h^L(k)$. However, the dCS interaction modifies the tensor equations of motion, generically leading to an amplitude birefringence, $P_h^R(k) \neq P_h^L(k)$. 

This parity violation is observationally quantified by the chirality parameter
\begin{equation}
    \chi(k) = \frac{P_h^R(k)-P_h^L(k)}{P_h^R(k)+P_h^L(k)}.
    \label{eq:chirality}
\end{equation}
The strict prediction of the present framework is a non-vanishing chirality ($\chi(k) \neq 0$). Detection of such a signature in the stochastic gravitational-wave background or via cross-correlations in the cosmic microwave background (e.g., non-zero $TB$ and $EB$ spectra) would constitute direct evidence for parity violation in the primordial gravitational sector.

\subsection{Propagation Effects: Birefringence and Damping}
\noindent 
The chiral interactions affect not only the production of gravitational waves but also their propagation through the post-inflationary universe. The dCS interaction dynamically induces different phase velocities for the two helicity states, $\omega_R(k) \neq \omega_L(k)$ \cite{Liu:2025ifb}. This phenomenon, known as gravitational-wave birefringence, causes the two polarizations to accumulate different phases over cosmological distances. The resulting birefringence angle scales with the integrated history of the condensate's chiral phase, offering a cumulative observational signature that can be searched for in future interferometric data.

Furthermore, the underlying fermionic dynamics induce gravitational-wave damping. As demonstrated in Ref.~\cite{Loverde:2022wih}, the propagation of tensor modes through a medium of self-interacting fermions is modified by the generation of anisotropic stress. In our framework, the exact same torsion-induced four-fermion interactions responsible for the initial condensation naturally provide this scattering cross-section. The GW damping rate is therefore intrinsically tied to the fundamental parameters of the model, specifically the cutoff scale $\Lambda$ and the critical density threshold $n_{A,\rm crit}$. Consequently, the precise shape of the damped GW spectrum encapsulates the microscopic thermodynamics of the density-driven waterfall.

\subsection{The Multi-Messenger Correlation: PBHs and Chiral GWs}
\noindent 
Crucially, parity violation in this model is not an ad hoc feature introduced via spectator fields; it is an unavoidable consequence of the same microscopic ECH dynamics that execute the waterfall transition. Torsion induces the four-fermion interaction, which drives both the inflationary condensate and the dCS gravity term. 

Because both sectors originate from this single unified mechanism, the most profound observational prediction of this framework is the strict theoretical correlation between the primordial black hole abundance and the gravitational-wave chirality:
\begin{equation}
    f_{\rm PBH} \quad \Longleftrightarrow \quad \chi(k).
\end{equation}
The same tachyonic fragmentation efficiency that shatters the condensate into Q-balls (yielding a larger $f_{\rm PBH}$) simultaneously excites large spatial gradients in the chiral phase, strongly sourcing the parity-violating tensor sector. 

Therefore, the framework predicts a unified multi-messenger signal: an extended PBH dark matter mass function, highly correlated with a chiral, birefringent, and structurally damped gravitational-wave background. Future observatories will have the capability to simultaneously constrain these observables, providing a direct and rigorous test of whether dark matter and cosmic inflation share a common, torsion-driven fermionic origin.

\section{Discussion and Outlook}
\label{sec:discussion}

\noindent 
Inflation remains one of the most successful paradigms in modern cosmology, yet its microscopic origins consistently elude standard particle physics. Most conventional models rely on the introduction of ad hoc fundamental scalar fields whose connections to known quantum matter remain tenuous. This theoretical gap strongly motivates the search for mechanisms in which inflation emerges dynamically from fundamental, pre-existing degrees of freedom.

In this work, we have developed a comprehensive framework in which the inflationary epoch originates entirely from fermion condensates, driven by torsion-induced four-fermion interactions within Einstein-Cartan-Holst (ECH) gravity \cite{Mercuri:2006um,Alexander:2014eva,Alexander:2014uaa,Freidel:2005sn}. The resulting composite states naturally play the role of the inflaton, circumventing the need for elementary scalars beyond those generated by the geometric self-interactions of the fermionic sector.

The central conceptual advance of this paper is the identification of a fundamentally new mechanism for terminating inflation: the density-driven waterfall. Unlike conventional hybrid inflation \cite{Linde:1993cn}, which exits geometrically when the rolling inflaton crosses a critical threshold in field space, the instability governed here is intrinsically thermodynamic. As continuous particle production proceeds during the inflationary expansion, the axial charge density $\langle n_A \rangle$ strictly increases, continuously modifying the phase space of the finite-density gap equation. Once this density breaches a critical topological threshold ($n_{A,\rm crit}$), the free energy cost of maintaining the broken-symmetry state exceeds that of a free Fermi gas. The macroscopic condensate instantaneously melts ($A \to 0$), terminating the inflationary phase not through a spatial boundary, but through the destruction of the inflaton medium itself.

This thermodynamic exit drastically departs from conventional cosmological paradigms by unifying multiple stages of early-universe evolution into a single microscopic origin. Phenomena traditionally treated as independent, decoupled sectors --- generating inflation, orchestrating the exit, executing reheating, and seeding dark matter --- all arise deterministically from the same composite sector. 

A profound consequence of the density-driven waterfall is the natural realization of instant preheating. Because the waterfall field is a collective excitation of fermions, its tachyonic collapse directly deposits the stored vacuum energy back into the underlying fermionic plasma via intrinsic $\mathcal{O}(1)$ Yukawa couplings. This circumvents the need for finely tuned mediator fields or external decay channels. Concurrently, the violently negative curvature of the effective potential during this transition shatters the condensate. This fragmentation deposits energy into stable, non-topological solitons (Q-balls), which emerge dynamically from the ECH vacuum. As these charge-conserving lumps cluster, they serve as the direct gravitational seeds for primordial black holes (PBHs) \cite{Coleman:1985ki,Kusenko:1997si,Cotner:2016dhw,Cotner:2016cvr,Cotner:2017tir,Cotner:2019ykd}, linking the inflationary exit directly to the astrophysics of compact objects.

Furthermore, this framework yields a highly distinctive, multi-messenger observational footprint. The fundamental torsion interactions responsible for fermion condensation are intrinsically chiral. Integrating out these fermions generates a dynamical Chern-Simons (dCS) coupling in the gravitational sector \cite{Alexander:2022cow}. Consequently, the exact same dynamics that seed PBHs simultaneously excite parity-violating signatures in the tensor metric perturbations \cite{Loverde:2022wih,Liu:2025ifb}. 

The characteristic prediction of our model is therefore a strictly correlated multi-messenger signal: an extended PBH mass function inextricably linked to a chiral, birefringent, and structurally damped stochastic gravitational-wave background. The detection of such correlations --- for instance, non-zero parity-violating $TB$ or $EB$ spectra in the cosmic microwave background alongside a measured PBH abundance --- would provide direct empirical evidence for this mechanism.

Several theoretical and phenomenological avenues remain open for future investigation. First, a fully non-linear 3D lattice simulation of the finite-density waterfall transition is required to accurately resolve the stochastic fragmentation dynamics. Such simulations will definitively map the Q-ball charge distribution to the terminal PBH mass function. Second, a dedicated computation of the chiral tensor power spectra, including the exact birefringence angle and GW damping coefficients, is necessary to translate our microscopic parameters into concrete templates for next-generation interferometers. Finally, the deep connection between fermionic many-body physics and dynamical Chern-Simons gravity suggests implications that extend well beyond inflation, potentially influencing late-time dark energy phenomenology and the broader landscape of parity-violating astrophysics.

Ultimately, this framework suggests a paradigm shift. Spacetime torsion induces chiral four-fermion interactions, which trigger macroscopic fermion condensation, driving inflation. The thermodynamic accumulation of charge forces a density-driven waterfall, resulting in instant preheating, the formation of Q-balls, gravitational collapse into PBHs, and the simultaneous radiation of chiral gravitational waves. Viewed from this perspective, inflation, reheating, dark matter formation, and GW parity violation are not disjoint cosmological epochs or independent theoretical sectors. Rather, they are sequential, unavoidable manifestations of a single torsion-induced fermionic phase. The density-driven waterfall mechanism thus provides a rigorous, unified bridge between quantum matter, spacetime geometry, and multi-messenger cosmology.

\appendix

\section{Derivation of the Effective Condensate Action}
\label{app:effective_action}

\noindent 
In this appendix, we detail the derivation of the low-energy effective action for the fermion condensate. We adapt the standard Nambu--Jona-Lasinio (NJL) formalism to a curved spacetime background, explicitly extracting the leading-order curvature corrections that couple to the composite scalar fields \cite{Inagaki:1993ya,PhysRevD.29.1584,Addazi:2017qus}.

\subsection{Bosonization of the Four-Fermion Interaction}
\noindent 
We begin with the effective four-fermion interaction derived from integrating out the non-dynamical torsion in the Einstein--Cartan--Holst theory, i.e. 
\begin{equation}
    \mathcal{L}_{\rm int} = \frac{1}{2M_\star^2} \left[ (\bar{\psi}\psi)^2 - (\bar{\psi}\gamma^5\psi)^2 \right].
\end{equation}
To linearize this interaction, we introduce a scalar auxiliary field $\Sigma$ and a pseudoscalar auxiliary field $\Pi$ via a Hubbard--Stratonovich transformation. The resulting bosonized action takes the form
\begin{equation}
    \begin{split}
        S_{\Sigma,\Pi} = \int d^4x\,\sqrt{-g} \Big[ &\bar{\psi}i\gamma^\mu\nabla_\mu\psi - \bar{\psi}(\Sigma+i\gamma^5\Pi)\psi \\
        &- \frac{M_\star^2}{2}(\Sigma^2+\Pi^2) \Big].
    \end{split}
\end{equation}
By defining the radial gap field $A^2 = \Sigma^2+\Pi^2$ and the chiral matrix field $\mathbf{A} = \Sigma+i\gamma^5\Pi$, the fermionic action becomes strictly bilinear in $\psi$, allowing the path integral over the fermion fields to be evaluated exactly.

\subsection{The One-Loop Effective Potential}
\noindent 
Integrating out the fundamental fermions yields the generating functional $Z = \int \mathcal{D}\Sigma \mathcal{D}\Pi \, e^{iS_{\rm eff}}$, where the one-loop quantum effective action is given by
\begin{equation}
    S_{\rm eff} = -\int d^4x\,\sqrt{-g} \frac{M_\star^2}{2} A^2 - i {\rm Tr} \ln \left( i\gamma^\mu\nabla_\mu - \mathbf{A} \right).
\end{equation}
Assuming a slowly varying background field, the effective potential of the condensate is defined by extracting the non-derivative terms from the action, namely
\begin{equation}
    V_{\rm eff}(A) = \frac{M_\star^2}{2} A^2 + i {\rm Tr} \ln \left( i\gamma^\mu\nabla_\mu - A \right).
    \label{eq:Vdet}
\end{equation}

To evaluate the functional determinant in curved spacetime, we express the trace logarithm as an integral over the fermionic Green's function $S(x,x';A) = (i\gamma^\mu\nabla_\mu - A)^{-1}$, i.e. 
\begin{equation}
    {\rm Tr}\ln \hat D = {\rm Tr} \int_0^A dA' \,S(x,x;A'),
\end{equation}
where the Dirac operator is $\hat D = i\gamma^\mu\nabla_\mu - A$. Following standard techniques \cite{PhysRevD.29.1584,Inagaki:1993ya}, we introduce an auxiliary function $G$ such that $S = (i\gamma^\mu\nabla_\mu + A)G$. This auxiliary Green's function consequently satisfies the Klein-Gordon-like equation, namely
\begin{equation}
    \left( \nabla^2 - A^2 \right) G = \delta(x,x').
\end{equation}

\subsection{Riemann Normal Coordinate Expansion}
\noindent 
To systematically solve for $G(x,x';A)$ in a curved background, we employ the Riemann normal coordinate (RNC) expansion centered around the point of interest \cite{PhysRevD.29.1584}. To first order in the spacetime curvature, the metric and its determinant expand as:
\begin{align}
    g_{\mu\nu} &= \eta_{\mu\nu} - \frac{1}{3} R_{\mu\rho\nu\sigma} y^\rho y^\sigma +\cdots , \\
    \sqrt{-g} &= 1 + \frac{1}{6} R_{\mu\nu} y^\mu y^\nu +\cdots .
\end{align}
Transforming to local momentum space, the perturbed Green's function takes the asymptotic form
\begin{equation}
    \label{eq:Gexp}
    \tilde G(q,A) = \frac{1}{q^2-A^2} - \frac{R}{12(q^2-A^2)^2} + \frac{2R_{\mu\nu}q^\mu q^\nu}{3(q^2-A^2)^3} +\cdots .
\end{equation}
Substituting Eq.~\eqref{eq:Gexp} back into Eq.~\eqref{eq:Vdet} allows us to cleanly separate the flat-spacetime contribution from the leading-order geometric curvature corrections.

\subsection{Final Potential and Critical Coupling}
\noindent 
Performing the Wick rotation to Euclidean space and integrating the momentum explicitly up to a physical ultraviolet cutoff scale $\Lambda$, we obtain the fully dressed effective potential
\begin{equation}
    \label{eq:appendixpotential}
    \begin{split}
        V_{\rm eff}(A) &= V_0 + \frac{M_\star^2}{2}A^2 \\
        &\quad - \frac{1}{4\pi^2} \Biggl[ A^2\Lambda^2 + \Lambda^4 \ln\left(1+\frac{A^2}{\Lambda^2}\right) \\
        &\quad \qquad \qquad - A^4 \ln\left(1+\frac{\Lambda^2}{A^2}\right) \Biggr] \\
        &\quad - \frac{R}{96\pi^2} \Biggl[ -A^2 \ln\left(1+\frac{\Lambda^2}{A^2}\right) + \frac{\Lambda^2 A^2}{\Lambda^2+A^2} \Biggr],
    \end{split}
\end{equation}
where $\mathcal{O}(R^2)$ terms have been systematically truncated as they are highly subdominant during the inflationary slow-roll phase ($H \ll \Lambda$). Equation~\eqref{eq:appendixpotential} serves as the foundational potential utilized throughout the main text.

Finally, the critical coupling separating the symmetric Fermi gas from the symmetry-broken condensed phase is identified by expanding the effective potential near the origin ($A \to 0$). Isolating the quadratic mass term yields
\begin{equation}
    V_{\rm eff}(A) \simeq \frac{1}{2} \left( M_\star^2 - \frac{\Lambda^2}{2\pi^2} \right) A^2 + \mathcal{O}(A^4).
\end{equation}
The dynamical phase transition therefore occurs at the critical coupling threshold, i.e. 
\begin{equation}
    M_{\star,c}^2 = \frac{\Lambda^2}{2\pi^2}.
\end{equation}
For interactions stronger than this critical value ($M_\star^2 < M_{\star,c}^2$), the origin becomes unconditionally tachyonic, driving the formation of the macroscopic condensate $A_{\rm vac} \neq 0$ that governs the inflationary dynamics modeled in this work.

\section{Covariant Kinetic Terms and Canonical Normalization}
\label{app:kinetic}

\noindent 
In this appendix, we derive the kinetic terms for the composite condensate fields and establish the canonical normalization required for the inflationary slow-roll analysis. The calculation employs the covariant heat-kernel expansion adapted for fermionic effective actions in curved spacetime \cite{PhysRevD.29.1584,Inagaki:1993ya}.

\subsection{The Squared Dirac Operator and Heat-Kernel Expansion}
\noindent 
The one-loop effective action derived in Appendix~\ref{app:effective_action} is governed by the functional determinant $S_{\rm 1-loop} = -i\,{\rm Tr}\ln(i\gamma^\mu\nabla_\mu - \mathbf{A})$, where $\mathbf{A} = \Sigma + i\gamma^5\Pi$. To construct a Laplace-type operator amenable to the heat-kernel expansion, we define the non-Hermitian Dirac operator $\hat D = i\gamma^\mu\nabla_\mu - \mathbf{A}$ and square it to form $\hat H = \hat D^\dagger \hat D$. 

Applying the Lichnerowicz identity for spinors, $\slashed{\nabla}^{\,2} = \nabla^2 - \frac{1}{4} R$, the squared operator takes the strictly second-order form $\hat H = -\nabla^2 + \mathbf{E}(x)$, where the endomorphism matrix $\mathbf{E}$ is given by
\begin{equation}
    \mathbf{E} = A^2 - \frac{1}{4}R - i\gamma^\mu \left( \partial_\mu\Sigma + i\gamma^5\partial_\mu\Pi \right).
    \label{eq:Eoperator}
\end{equation}
The derivative terms contained within $\mathbf{E}(x)$ are the direct source of the dynamically generated kinetic energy for the composite condensate fields.

We express the effective action using the Schwinger proper-time representation, namely
\begin{equation}
    S_{\rm eff} = -\frac{1}{2} {\rm Tr} \int_{1/\Lambda^2}^\infty \frac{ds}{s} e^{-s\hat H}.
\end{equation}
The heat kernel admits the standard asymptotic expansion ${\rm Tr}\,e^{-s\hat H} = (4\pi s)^{-2} \sum_{n=0}^{\infty} s^n a_n(x)$, where $a_n$ are the Seeley-DeWitt coefficients. The emergent kinetic terms arise from the logarithmically divergent contribution captured by the $a_2$ coefficient \cite{PhysRevD.29.1584}, specifically the term $a_2 \supset \frac{1}{2} {\rm tr}(\mathbf{E}^2)$. 

Substituting Eq.~\eqref{eq:Eoperator} and evaluating the Dirac trace (noting that cross terms carrying a single $\gamma^5$ vanish identically), we find:
\begin{equation}
    {\rm tr}(\mathbf{E}^2) = -4 g^{\mu\nu}\partial_\mu\Sigma\partial_\nu\Sigma - 4 g^{\mu\nu}\partial_\mu\Pi\partial_\nu\Pi + \cdots .
\end{equation}

\subsection{Induced Kinetic Terms and Canonical Fields}
\noindent 
Integrating over the proper time $s$ with the physical ultraviolet cutoff $\Lambda$, one obtains that the emergent kinetic Lagrangian for the condensate takes the form
\begin{equation}
    \mathcal{L}_{\rm kin} = Z_A g^{\mu\nu} \left( \partial_\mu\Sigma\partial_\nu\Sigma + \partial_\mu\Pi\partial_\nu\Pi \right),
    \label{eq:kineticlag}
\end{equation}
where the wavefunction renormalization factor $Z_A$ is defined via the incomplete Gamma function, i.e. 
\begin{equation}
    Z_A = \frac{1}{16\pi^2} \int_{1/\Lambda^2}^{\infty} \frac{ds}{s} e^{-sA^2} = \frac{1}{16\pi^2} \Gamma\left(0, \frac{A^2}{\Lambda^2}\right).
    \label{eq:ZAproper}
\end{equation}
During the inflationary regime, the condensate amplitude is deeply sub-Planckian ($A^2 \ll \Lambda^2$). In this limit, the incomplete Gamma function expands as $\Gamma(0,x) = -\gamma_E - \ln x + \mathcal{O}(x)$, yielding a positive-definite kinetic coefficient that varies only logarithmically along the inflationary trajectory, namely
\begin{equation}
    Z_A \simeq \frac{1}{16\pi^2} \left[ -\gamma_E - \ln\left(\frac{A^2}{\Lambda^2}\right) \right].
\end{equation}

Since $Z_A$ is field-dependent, the bare condensate amplitudes $A$ and $B$ (from the bipartite sector) are not canonical variables. To rigorously evaluate the slow-roll dynamics, we must map the system to canonically normalized fields:
\begin{equation}
    \Phi_A = \int^A dA' \sqrt{2Z_A(A')}, \quad \Phi_B = \int^B dB' \sqrt{2Z_B(B')}.
    \label{eq:canonicalAB}
\end{equation}
In terms of these physical variables, the kinetic sector diagonalizes to $\mathcal{L}_{\rm kin} = \frac{1}{2}(\partial\Phi_A)^2 + \frac{1}{2}(\partial\Phi_B)^2$. The inflationary slow-roll parameters evaluated in the main text are strictly defined with respect to these canonical fields:
\begin{equation}
    \epsilon = \frac{M_{\rm pl}^2}{2} \left( \frac{\partial_{\Phi_B}V}{V} \right)^2, \qquad \eta = M_{\rm pl}^2 \frac{\partial_{\Phi_B}^2 V}{V}.
\end{equation}
This derivation assumes the scale hierarchy $H^2 \ll \Lambda^2$, ensuring that higher-order curvature corrections in the heat-kernel expansion remain deeply suppressed, thereby validating the effective single-field slow-roll description.

\section{Finite-Density Effective Potential and Gap Equation}
\label{app:finite_density}

\noindent 
The density-driven waterfall mechanism central to this work is a strict consequence of finite-density thermodynamics within the strongly coupled fermion plasma. This appendix details the derivation of the finite-density effective potential, the modified gap equation, and the axial-density constraint, extending the standard NJL formalism to incorporate continuous chiral charge production \cite{Inagaki:1993ya}.

\subsection{Chemical Potential Deformation and the Euclidean Spectrum}
\noindent 
To formally model the accumulation of axial charge during the inflationary epoch, we deform the Lagrangian by introducing a chiral chemical potential $\mu$, i.e. 
\begin{equation}
    \mathcal{L}_\mu = - \mu \bar{\psi} \gamma^0 \gamma^5 \psi.
    \label{eq:appchemical}
\end{equation}
The deformed Dirac operator becomes $\mathcal{D} = i\gamma^\mu\nabla_\mu - A - \mu\gamma^0\gamma^5$. The one-loop quantum effective potential is structurally defined by the determinant $V(A,\mu) = A^2/(2\lambda) - i {\rm Tr}\ln\mathcal{D}$. Since the chemical potential explicitly breaks Lorentz invariance by selecting a preferred cosmic rest frame, the Riemann normal-coordinate expansion utilized in Appendix~\ref{app:effective_action} is suboptimal. Instead, we evaluate the determinant directly in momentum space.

The presence of the chiral chemical potential splits the fermionic spectrum into two distinct helicity branches, characterized by the modified dispersion relations
\begin{equation}
    E_\pm = \sqrt{ (|\vec{q}|\pm\mu)^2 + A^2 }.
    \label{eq:Epm}
\end{equation}
Performing a Wick rotation to Euclidean space ($q_0 = i\omega$), the functional trace transforms into an explicit momentum-space integral. Imposing a cylindrical cutoff scheme consistent with the background symmetries, $W_q = \sqrt{\Lambda^2-q^2}$, the Euclidean effective potential evaluates to
\begin{equation}
    \label{eq:VEuclidean}
    \begin{split}
        V(A,\mu) &= \frac{A^2}{2\lambda} - \frac{1}{4\pi^3} \int_0^\Lambda dq\,q^2 \int_{-W_q}^{W_q} d\omega \\
        &\quad \times \ln \Big[ (\omega^2+E_+^2) (\omega^2+E_-^2) \Big].
    \end{split}
\end{equation}
This function completely encapsulates the finite-density vacuum structure of the condensate.

\subsection{The Chiral Gap Equation and Axial Density}
\noindent 
The physical configuration of the macroscopic condensate is determined by the stationary points of the effective potential, $\partial V/\partial A = 0$. Executing the integration over the Euclidean energy $\omega$ yields the finite-density gap equation for the broken phase ($A \neq 0$), namely
\begin{equation}
    \label{eq:appendixgap}
    \begin{split}
        \frac{1}{\lambda} &= \frac{1}{\pi^3} \int_0^\Lambda dq\,q^2 \Biggl[ \frac{1}{E_+} \tan^{-1} \left( \frac{W_q}{E_+} \right) \\
        &\quad + \frac{1}{E_-} \tan^{-1} \left( \frac{W_q}{E_-} \right) \Biggr].
    \end{split}
\end{equation}
Concurrently, the macroscopic observable thermodynamically conjugate to the chemical potential is the axial number density, $n_A = - \partial V / \partial\mu$. Differentiating Eq.~\eqref{eq:VEuclidean} provides the state equation for the chiral plasma, i.e. 
\begin{equation}
    \label{eq:appendixdensity}
    \begin{split}
        n_A &= \frac{1}{\pi^3} \int_0^\Lambda dq\,q^2 \Biggl[ \frac{q+\mu}{E_+} \tan^{-1} \left( \frac{W_q}{E_+} \right) \\
        &\quad - \frac{q-\mu}{E_-} \tan^{-1} \left( \frac{W_q}{E_-} \right) \Biggr].
    \end{split}
\end{equation}
Equations~\eqref{eq:appendixgap} and \eqref{eq:appendixdensity} constitute a strictly closed thermodynamic system, defining a one-dimensional equilibrium vacuum manifold embedded within the parameter space $(A,\mu)$. 

\subsection{Topological Bifurcation and the Dynamical Exit}
\noindent 
The fundamental mechanism driving the termination of inflation lies in the geometry of this vacuum manifold. Because the phase space is strictly bounded by the cutoff $\Lambda$, the manifold does not extend to arbitrarily large axial densities. It terminates abruptly at a turning point where $dn_A/d\mu = 0$, defining a strict critical threshold, $n_{A,\rm crit}$. 

For densities below this threshold ($n_A < n_{A,\rm crit}$), the gap equation admits a stable, symmetry-broken solution, sustaining the slow-roll phase. However, beyond this critical limit ($n_A > n_{A,\rm crit}$), the energetic cost of maintaining the condensate outpaces the phase-space availability of the Fermi sea. The mathematical solution to the gap equation vanishes entirely, enforcing $A_{\rm vac} = 0$.

During the inflationary expansion, anomalous particle production continuously pumps axial charge into the background \cite{Adshead:2015kza,Chen:2018xck,Adshead:2018oaa,Hook:2019zxa,Sou:2021juh,Chen:2023txq}. Governed by the anomaly equation, $\dot n_A + 3Hn_A = \Gamma_A$, the system is forcefully driven toward this topological boundary ($n_A(t) \to n_{A,\rm crit}$). The density-driven waterfall is therefore not an externally imposed field-space coordinate, but a dynamically inevitable thermodynamic instability, providing a robust, microscopic realization of the inflationary exit.

\section{Linear Instability Analysis and Q-Ball Formation}
\label{app:qballs}

\noindent 
In this appendix, we rigorously derive the linear instability conditions responsible for the fragmentation of the macroscopic condensate and the subsequent formation of Q-balls. The analysis adapts the standard treatment of non-topological soliton formation in continuous phase transitions \cite{Coleman:1985ki,Kusenko:1997si,Kusenko:1997vp,Dvali:1997qv,Kasuya:2000wx,Kasuya:2001hg,Multamaki:1999an,Cotner:2019ykd} to the specific composite dynamics of the ECH fermion condensate sector.

\subsection{Complex Condensate Dynamics and Perturbations}
\noindent 
To analyze the post-waterfall symmetry breaking, we parameterize the low-energy condensate field in terms of its radial amplitude $\rho$ and chiral phase $\theta$, i.e. 
\begin{equation}
    \varphi = \frac{1}{\sqrt{2}} \rho\,e^{i\theta}.
\end{equation}
The effective action governing this sector is given by $S = \int d^4x \sqrt{-g} \left[ |\partial_\mu\varphi|^2 - V(\rho) \right]$. Assuming a spatially flat Friedmann--Lema\^{i}tre--Robertson--Walker (FLRW) background, $ds^2 = -dt^2 + a^2(t)d\vec{x}^2$, the fully coupled nonlinear equations of motion for the real scalar components are
\begin{align}
    \ddot{\rho} + 3H\dot{\rho} - \frac{\nabla^2\rho}{a^2} - \rho\dot{\theta}^2 + \frac{\partial V}{\partial\rho} &= 0, \label{eq:rhoeom} \\
    \ddot{\theta} + 3H\dot{\theta} - \frac{\nabla^2\theta}{a^2} + 2\frac{\dot{\rho}}{\rho}\dot{\theta} - \frac{2}{a^2\rho}\nabla\rho\cdot\nabla\theta &= 0. \label{eq:thetaeom}
\end{align}
These equations dictate the highly dynamic relaxation of the condensate immediately following the density-driven waterfall.

To study the onset of spatial fragmentation, we decompose the fields into a homogeneous rolling background and small spatial fluctuations: $\rho(t,\vec{x}) = \rho_0(t) + \delta\rho(t,\vec{x})$ and $\theta(t,\vec{x}) = \theta_0(t) + \delta\theta(t,\vec{x})$. Expanding the perturbations into Fourier momentum modes, we follow the standard instability ansatz \cite{Kusenko:1997si,Kasuya:2000wx,Kasuya:2001hg} and seek solutions of the form $\delta X_k \propto e^{\alpha t - i\vec{k}\cdot\vec{x}}$, where $\alpha$ represents the dynamical growth exponent. A specific momentum mode becomes violently unstable against fragmentation whenever ${\rm Re}(\alpha) > 0$.

\subsection{The Dispersion Relation and Tachyonic Instability}
\noindent 
Linearizing Eqs.~\eqref{eq:rhoeom} and \eqref{eq:thetaeom} around the homogeneous background $(\rho_0, \theta_0)$ yields a coupled system for the perturbative modes. Defining the background angular velocity as $\omega \equiv \dot{\theta}_0$, the combination of these linearized equations generates the characteristic dispersion relation for the growth exponent $\alpha$:
\begin{equation}
    \label{eq:dispersion}
    \begin{split}
        &4\omega^2 \alpha \left( \alpha - \frac{\dot{\rho}_0}{\rho_0} \right) \\
        &+ \left( \alpha^2 + 3H\alpha + \frac{k^2}{a^2} + 2\frac{\dot{\rho}_0}{\rho_0}\alpha \right) \\
        &\times \left( \alpha^2 + 3H\alpha + \frac{k^2}{a^2} + V''(\rho_0) - \omega^2 \right) = 0.
    \end{split}
\end{equation}
This characteristic equation completely determines the linear stability properties of the rolling condensate. 

To identify the phase-space boundary of the instability band, we set the marginal growth condition $\alpha = 0$. Assuming the expansion rate $H$ is subleading during the rapid waterfall transition, Eq.~\eqref{eq:dispersion} simplifies significantly, yielding the boundaries
\begin{equation}
    \frac{k^2}{a^2} \left[ \frac{k^2}{a^2} + V''(\rho_0) - \omega^2 \right] = 0.
\end{equation}
The exponentially growing, unstable modes are therefore confined strictly to the momentum band, namely
\begin{equation}
    0 < \frac{k^2}{a^2} < \omega^2 - V''(\rho_0).
    \label{eq:instband}
\end{equation}
Consequently, a robust physical instability occurs whenever $V''(\rho_0) < \omega^2$. 

Crucially, the effective potential generated by the fermionic determinant (derived in Appendix~\ref{app:effective_action}) inherently exhibits a broad, flattened structure at large field values, leading directly to regions where the curvature is strictly negative, $V''(\rho_0) < 0$. During the post-waterfall relaxation phase, the macroscopic field is forced to traverse this tachyonic region. The instability condition in Eq.~\eqref{eq:instband} is therefore unconditionally satisfied. The density-driven waterfall mechanism dynamically and unavoidably creates the precise thermodynamic environment required to shatter the homogeneous condensate.

\subsection{Stationary Q-Ball Solutions and Scaling Relations}
\noindent 
The ultimate endpoint of this tachyonic fragmentation is the formation of localized, stable, non-topological solitons. As the growing modes become strongly nonlinear, the field configuration settles into localized stationary solutions of the form $\varphi(t,\vec{x}) = \frac{1}{\sqrt{2}}\rho(r)e^{i\omega t}$. 

The radial profile $\rho(r)$ is determined by minimizing the energy functional at a fixed global $U(1)_A$ charge $Q = \omega \int d^3x\,\rho^2$. The corresponding energy functional evaluates to:
\begin{equation}
    E_Q = \int d^3x \left[ \frac{1}{2}(\nabla\rho)^2 + V(\rho) + \frac{1}{2}\omega^2\rho^2 \right].
\end{equation}
Variation of this functional yields the fundamental profile equation:
\begin{equation}
    \rho'' + \frac{2}{r}\rho' + \omega^2\rho - \frac{dV}{d\rho} = 0.
    \label{eq:qprofileapp}
\end{equation}
Because the ECH effective potential asymptotes to a constant vacuum energy scale ($V \to V_0 \sim \Lambda^4$), the resulting solitons are "flat-potential" Q-balls. Minimization of the energy for this specific class of potentials enforces the characteristic phase-velocity scaling $\omega \propto \Lambda Q^{-1/4}$. 

Applying this relation to the integral observables generates the strict macroscopic scaling laws for the solitons:
\begin{equation}
    E_Q \propto \Lambda Q^{3/4}, \qquad R_Q \propto \frac{Q^{1/4}}{\Lambda}.
\end{equation}
These scaling relations, extensively employed in Sec.~\ref{sec:qballs}, guarantee that the energy-per-charge ratio ($E_Q/Q$) decreases monotonically with size, rendering large Q-balls kinematically stable against decay into the underlying fundamental fermions. 

\section{Primordial Black Hole Mass Function}
\label{app:pbh}

\noindent 
This appendix details the derivation of the primordial black hole (PBH) mass function generated by the stochastic clustering of the Q-ball population discussed in Sec.~\ref{sec:qballs}. We adapt the statistical framework developed in Refs.~\cite{Cotner:2016dhw,Cotner:2016cvr,Cotner:2017tir,Cotner:2019ykd} to the specific composite fermion-condensate dynamics characterizing the density-driven waterfall.

\subsection{Stochastic Clustering and Poisson Statistics}
\noindent 
The tachyonic fragmentation process generates a robust population of Q-balls, each carrying an approximately conserved axial charge $Q$ and a characteristic mass $M_Q = \Lambda Q^{3/4}$. Because the spatial phase of the tachyonic instability is inherently random, these Q-balls are distributed stochastically throughout the post-inflationary Universe. 

Consider a localized comoving volume $V$ containing exactly $N$ Q-balls. The total mass contained within this specific volume is $M = N\,M_Q$. Macroscopic density fluctuations arise entirely because the discrete number of Q-balls fluctuates from region to region. Following Refs.~\cite{Cotner:2016dhw,Cotner:2016cvr,Cotner:2017tir,Cotner:2019ykd}, we model this initial Q-ball spatial distribution as a purely Poissonian process, namely
\begin{equation}
    P(N|V) = \frac{(\bar n V)^N}{N!} e^{-\bar n V},
    \label{eq:poissonapp}
\end{equation}
where $\bar n$ denotes the average cosmological Q-ball number density. The mean expected number of solitons in the volume is $\langle N\rangle = \bar n V$, with an identical variance $\sigma_N^2 = \bar n V$. 

Consequently, the relative density fluctuations driven by this discrete clustering scale precisely as
\begin{equation}
    \frac{\delta\rho}{\rho} \sim \frac{\sigma_N}{\langle N \rangle} = \frac{1}{\sqrt{\bar n V}}.
\end{equation}
This inverse square-root scaling guarantees that rare, high-density topological regions will naturally and unavoidably emerge from the stochastic fragmentation of the condensate, providing the necessary overdense seeds for gravitational collapse.

\subsection{Cluster Mass Distribution and the Collapse Criterion}
\noindent 
To determine the cosmological abundance of these overdense seeds, we must compute the probability of finding a cluster with a total aggregate mass $M$ inside a given volume $V$. This is obtained by marginalizing over the discrete number distribution, i.e. 
\begin{equation}
    P(M|V) = \sum_N P(M|N) P(N|V).
\end{equation}
Assuming for analytical simplicity a sharply peaked charge distribution where the Q-balls are nearly identical, the mass-number relation is strictly a delta function, $P(M|N) = \delta(M-NM_Q)$. The continuous cluster mass distribution therefore becomes 
\begin{equation}
    P(M|V) = \sum_N \delta(M-NM_Q) \frac{(\bar nV)^N}{N!} e^{-\bar nV}.
    \label{eq:massdistribution}
\end{equation}
Equation~\eqref{eq:massdistribution} dictates the exact statistical frequency of overdense regions generated by the Q-ball population.

However, an overdense cluster only undergoes catastrophic collapse into a PBH if its physical scale is sufficiently compact. Specifically, the cluster's localization must fall within its own Schwarzschild radius, $R_{\rm BH} = 2M/M_{\rm pl}^2$. Following the established collapse formalism \cite{Kodama:1986ud,Cotner:2019ykd}, we encode this geometric threshold into a strict collapse probability filter, $B(M,V)$, which specifies the fraction of clusters of mass $M$ and volume $V$ that successfully satisfy the Schwarzschild criterion against internal velocity dispersions.

Integrating over the scale-invariant volume measure $dV/V^2$, the macroscopic, spatially averaged PBH energy density is given by
\begin{equation}
    \langle\rho_{\rm PBH}\rangle = \int \frac{dV}{V^2} \int dM\, P(M|V) B(M,V) M.
    \label{eq:rhopbh}
\end{equation}

\subsection{The PBH Mass Function and Dark Matter Fraction}
\noindent
The primary observable derived from this population is the differential PBH mass function, $d\langle\rho_{\rm PBH}\rangle/dM$. Substituting the mass distribution from Eq.~\eqref{eq:massdistribution} into Eq.~\eqref{eq:rhopbh}, we obtain the fundamental semi-analytical expression for the PBH spectrum, i.e. 
\begin{equation}
    \label{eq:pbhmassfunction}
    \begin{split}
        \frac{d\langle\rho_{\rm PBH}\rangle}{dM} &= \sum_N \int \frac{dV}{V^2} \, M \, B(M,V) \\
        &\quad \times \delta(M-NM_Q) \frac{(\bar nV)^N}{N!} e^{-\bar nV}.
    \end{split}
\end{equation}
To connect this absolute density to astrophysical observations, we express it as a differential fraction of the total cosmological dark matter density, $\rho_{\rm DM}$. The logarithmic differential fraction is defined as:
\begin{equation}
    \frac{df_{\rm PBH}}{d\ln M} = \frac{1}{\rho_{\rm DM}} \frac{d\langle\rho_{\rm PBH}\rangle}{d\ln M}.
\end{equation}
Integrating this spectrum yields the total PBH dark matter abundance, $f_{\rm PBH} = \int d\ln M \, (df_{\rm PBH}/d\ln M)$. 

The total abundance and the peak position of this mass function are intrinsically governed by the underlying condensate physics. The ECH inflationary framework introduces a natural ultraviolet scale, $\Lambda \sim 10^{-3}M_{\rm pl}$. Processed through the Q-ball mass scaling $M_Q = \Lambda Q^{3/4}$, this strongly dictates the characteristic mass scale of the resulting compact objects. However, due to the stochastic sum over $N$ in Eq.~\eqref{eq:pbhmassfunction}, the model generically predicts an extended, broad mass function rather than a monochromatic spike. This extended nature is phenomenologically critical for navigating tight, mass-dependent observational constraints from microlensing and Hawking evaporation \cite{Carr:2016drx,Sasaki:2016jop}.

Ultimately, this derivation rigorously demonstrates that the PBH population is not an ad hoc addition to the cosmological timeline. The mass function $df_{\rm PBH}/d\ln M$ acts as a direct, observable historical record of the density-driven waterfall mechanism. Measurements of the PBH spectrum can therefore be inverted to directly constrain the microscopic parameters (such as the cutoff $\Lambda$ and critical density $n_{A,\rm crit}$) that orchestrate the termination of the inflationary epoch.

\section{Parity-Violating Gravitational Waves from the Condensate Sector}
\label{app:dcs}

\noindent
This appendix derives the parity-violating tensor dynamics discussed in Sec.~\ref{sec:gws}, establishing the rigorous mathematical connection between the microscopic fermion condensate sector and macroscopic gravitational-wave observables. The derivation adopts the effective-field-theory framework developed for torsion-induced chiral gravity in Refs.~\cite{Alexander:2022cow,Liu:2025ifb,Loverde:2022wih}.

\subsection{Emergent Dynamical Chern-Simons Gravity}
\noindent
The torsion-induced fermionic interactions responsible for the inflationary condensation possess a nontrivial, intrinsic chiral structure. Upon integrating out these fermionic degrees of freedom to obtain the low-energy effective theory, the chiral anomaly inevitably generates parity-odd contributions to the gravitational action \cite{Alexander:2022cow}. This manifests as an emergent dynamical Chern-Simons (dCS) term, namely
\begin{equation}
    S_{\rm dCS} = \int d^4x \sqrt{-g} \, \frac{\vartheta}{4} \,{}^\ast RR,
    \label{eq:dcsaction}
\end{equation}
where ${}^\ast RR = \frac{1}{2} \epsilon^{\mu\nu\rho\sigma} R_{\mu\nu\alpha\beta} R_{\rho\sigma}{}^{\alpha\beta}$ denotes the gravitational Pontryagin density. The pseudoscalar field $\vartheta$ arises directly from the chiral phase of the condensate sector, inheriting its macroscopic dynamics from the underlying finite-density fermionic theory. The explicit parity violation in Eq.~\eqref{eq:dcsaction} fundamentally modifies the classical propagation of tensor metric perturbations.

\subsection{Chiral Tensor Perturbations and Modified Dynamics}
\noindent
To analyze the gravitational-wave signatures, we perturb the spatial metric around the flat FLRW background, $g_{ij} = a^2(t)(\delta_{ij} + h_{ij})$, enforcing the standard transverse-traceless gauge conditions ($\partial_i h_{ij} = 0, h_{ii} = 0$). The tensor field is decomposed into the circular polarization basis, which is
\begin{equation}
    h_{ij} = \sum_{\lambda=R,L} h_\lambda e^{(\lambda)}_{ij}.
\end{equation}
The polarization tensors $e^{(\lambda)}_{ij}$ are eigenstates of the curl operator, satisfying $i \epsilon_{ilm} \hat{k}_l e^{(\lambda)}_{mj} = \lambda e^{(\lambda)}_{ij}$, with helicity eigenvalues $\lambda = +1$ for right-handed (R) and $\lambda = -1$ for left-handed (L) modes.

In the presence of the dCS interaction, the equations of motion for the tensor perturbations acquire a helicity-dependent source term, i.e. 
\begin{equation}
    \ddot h_\lambda + 3H\dot h_\lambda + \frac{k^2}{a^2}h_\lambda = \lambda \Delta_{\rm CS} h_\lambda.
    \label{eq:tensoreom}
\end{equation}
The modification factor $\Delta_{\rm CS} \propto (k/a)\dot{\vartheta}$ encodes the parity-violating geometric correction \cite{Alexander:2022cow,Liu:2025ifb}. Equation~\eqref{eq:tensoreom} demonstrates that the left- and right-handed helicities no longer evolve symmetrically; the dynamical running of the condensate phase ($\dot{\vartheta} \neq 0$) breaks the degeneracy, amplifying one circular polarization while suppressing the other.

\subsection{Tensor Chirality, Birefringence, and Damping}
\noindent
This asymmetric evolutionary trajectory imprints permanent signatures on the observable tensor power spectra, defined via $\langle h_\lambda(\vec{k}) h_{\lambda'}(\vec{k}') \rangle = (2\pi)^3 \delta_{\lambda\lambda'} \delta(\vec{k}+\vec{k}') P_h^\lambda(k)$. In standard parity-conserving General Relativity, $P_h^R = P_h^L$. In our condensate framework, the dCS interaction yields an amplitude birefringence ($P_h^R \neq P_h^L$), quantified by the dimensionless chirality parameter 
\begin{equation}
    \chi(k) = \frac{P_h^R(k)-P_h^L(k)}{P_h^R(k)+P_h^L(k)}.
    \label{eq:chiralityapp}
\end{equation}
A strictly non-zero chirality ($\chi \neq 0$) is a fundamental, falsifiable prediction of this model.

Beyond amplitude asymmetry, the dCS interaction modifies the propagation kinematics, inducing distinct dispersion relations for the two helicities, $\omega_R(k) \neq \omega_L(k)$. Over cosmological distances, this phase velocity differential generates gravitational-wave birefringence \cite{Liu:2025ifb}. The accumulated phase shift, $\Delta\Phi = \int dt \, (\omega_R - \omega_L)$, rotates the linear polarization plane of the stochastic background, providing a distinct interferometric observable.

Furthermore, the propagation of tensor modes through the dense fermionic plasma is subject to anisotropic stress generated by the same four-fermion interactions that drive the condensation. As shown in Ref.~\cite{Loverde:2022wih}, this introduces a macroscopic damping coefficient $\Gamma_{\rm GW}(M_\star,\Lambda,n_A)$ into the tensor equation, i.e. 
\begin{equation}
    \ddot h_\lambda + (3H+\Gamma_{\rm GW})\dot h_\lambda + \left(\frac{k^2}{a^2} - \lambda \Delta_{\rm CS}\right) h_\lambda = 0.
\end{equation}
Consequently, the precise shape and amplitude of the gravitational-wave spectrum carry direct information about the microscopic cutoff $\Lambda$ and the axial density $n_A$.

\subsection{Multi-Messenger Correlation with the Waterfall Mechanism}
\noindent
Crucially, the parity-violating sector in this framework is not an ad hoc addition; it is deterministically coupled to the inflationary exit. The density-driven waterfall mechanism strictly governs the temporal evolution of the chiral phase, directly supplying the $\dot{\vartheta}$ term that seeds $\chi(k)$.

Because the same density-driven instability simultaneously triggers the formation of Q-balls (which subsequently cluster into primordial black holes), the framework predicts a strict, unified correlation between the dark matter and gravitational-wave sectors:
\begin{equation}
    f_{\rm PBH} \quad \Longleftrightarrow \quad \chi(k).
\end{equation}
An environment with highly efficient tachyonic fragmentation not only elevates the PBH abundance ($f_{\rm PBH}$), but also violently excites gradients in the condensate phase, maximizing the chiral GW signal. 

This model therefore provides a rich, testable web of observables. Future cosmic microwave background (CMB) measurements of parity-violating $TB$ and $EB$ correlations, combined with interferometric detections of chiral tensor spectra, GW birefringence, structural damping, and PBH abundances, will not merely test isolated parameters. Rather, their simultaneous observation would provide definitive, multi-messenger validation of the torsion-induced, density-driven fermion condensate mechanism in the early Universe.

\bibliographystyle{apsrev4-2}
\bibliography{reference}
\end{document}